\documentclass[12pt, preprint]{aastex}

\begin{document}

\title{{\it RHESSI} Observation of Chromospheric Evaporation}

\author{Wei Liu\altaffilmark{1}, Siming Liu\altaffilmark{2}, Yan Wei Jiang\altaffilmark{1}, 
and Vah\'{e} Petrosian\altaffilmark{3}}

\affil{Center for Space Science and Astrophysics, Stanford University, Stanford, California 94305}

\altaffiltext{1}{Also Department of Physics; weiliu@sun.stanford.edu, arjiang@stanford.edu} 
\altaffiltext{2}{Current address: Los Alamos National Laboratory, Los Alamos, NM 87545; liusm@lanl.gov}
\altaffiltext{3}{Also Departments of Physics and Applied Physics; vahe@astronomy.stanford.edu}

\begin{abstract}

We present analyses of the spatial and spectral evolution of hard X-ray emission
observed by {\it RHESSI} during the impulsive phase of an M1.7 flare on 2003 November 13. 
In general, as expected, the loop top (LT) source dominates at low energies while 
the footpoint (FP) sources dominate the high energy emission. At intermediate 
energies, both the LT and FPs may be seen, but during certain intervals emission from the 
legs of the loop dominates, in contrast to the commonly observed LT and FP emission. 
The hard X-ray emission tends to rise above the FPs and eventually merge into
a single LT source. This evolution starts first at low energies and proceeds to higher energies.
The spectrum of the resultant LT source becomes more and more dominated 
by a thermal component with an increasing emission measure as the flare proceeds. 
The soft and hard X-rays show a Neupert-type behavior.
With a non-thermal bremsstrahlung model the brightness profile along the loop
is used to determine the density profile and its evolution, which reveals a 
gradual increase of the gas density in the loop.
These results are evidence for chromospheric evaporation and are consistent 
with the qualitative features of hydrodynamic simulations of this phenomenon. 
However, some observed source morphology and its evolution cannot be
accounted for by previous simulations. Therefore simulations 
with more realistic physical conditions are required to explain 
the results and the particle acceleration and plasma heating processes.

\end{abstract}



\section{Introduction}
\label{intr}

Chromospheric evaporation was first suggested 
by \citet{Neup68} to explain the origin of the hot, dense, soft X-ray (SXR) emitting plasma 
confined in the coronal loops during solar flares. The basic scenario is as follows.  
Magnetic reconnection, believed to be the primary energy release mechanism, heats the plasma and 
accelerates particles high in the corona. The released energy is transported downward along the 
newly reconnected closed flaring loop by non-thermal particles and/or thermal conduction, heating the 
chromospheric material rapidly (at a rate faster than the radiative and conductive cooling rates) 
up to a temperature of $\sim$$10^7$ K.  The resulting overpressure drives a mass flow upward 
along the loop at a speed of a few hundreds of km s$^{-1}$, which fills the flaring loop with 
a hot plasma giving rise to the gradual evolution of SXR emission. This
process should also result in a derivative of the SXR light curve in its rising 
portion that closely matches the hard X-ray (HXR) light curve, which is called 
Neupert effect and is observed in some (but not all) flares \citep{Neup68, Huds91, Denn93, Denn03, Vero05}.

Hydrodynamic (HD) simulations of chromospheric evaporation have been carried out with an assumed 
energy transport mechanism (e.g., electron ``beam'' or conductive heating) \citep{Fish85a, 
Mari89, Gan95, Yoko01, Allr05} leading to various predictions on the UV-SXR spectral lines 
produced by the evaporated plasma, as well as the density and temperature profiles along 
the flaring loop. Most of the observational tests of these predictions rely on the blue-shifted 
components of SXR emission lines produced by the up-flowing plasma, first reported by 
\citet{Dosc80} and \citet{Feld80} using spectra obtained from the {\it P78-1} spacecraft. 
Similar observations were subsequently obtained from X-ray spectrometers on the Solar Maximum Mission
({\it SMM}; Antonucci et al. 1982, 1984), 	
the {\it Hinotori} spacecraft \citep{Wata90},
the {\it Yohkoh} spacecraft \citep{Wuls94}, and the Solar and Heliospheric Observatory 
({\it SoHO}; Brosius 2003; Brosius \& Philips 2004).  
\citet{Wuls94}, on the other hand, observed co-spatial SXR 
blue-shifts (up-flows) and H$_{\alpha}$ red-shifts (down-flows) as expected from HD simulations 
\citep{Fish85b}. 
A summary of relevant observations from {\it SMM} can be found in \citet{Anto99}.

All the aforementioned observations, however, were indirect evidence in the sense that the 
evaporation process was not imaged directly. Based on HD simulations, \citet{Pere93} derived the 
expected X-ray brightness profile across the evaporation front and suggested that {\it 
Yohkoh}/SXT or X-ray imagers with equivalent or better spatial and temporal resolution should be 
able to detect the front. Indeed, \citet{Silv97} found that the HXR and SXR sources of the 1994 
June 30 flare moved toward the loop top (LT) during the impulsive phase. Since the flare was
located near the center of the solar disk, 
they identified such motions as the horizontal counterpart of the 
line-of-sight motion revealed by the blue-shifted emission lines observed simultaneously by 
{\it Yohkoh}/BCS.

The Reuven Ramaty High Energy Solar Spectroscopic Imager ({\it RHESSI})
with its superior spatial, temporal, and spectral resolution \citep{Lin02} 
provides us with opportunities to study the chromospheric 
evaporation process in unprecedented detail.  We report in this paper our analyses of the 
spatial and spectral evolution of a simple flare on 2003 November 13 with excellent {\it RHESSI} 
coverage. Because the flare occurred near the solar limb it presented minimum projection effects and 
a well-defined loop geometry that allows direct imaging of the HXR brightness profile along the 
loop.  The observations and data analyses are presented in \S \ref{obs}, followed by a derivation of 
the evolution of the density profile along the flaring loop in \S \ref{dens}. 
We summarize the major findings of 
this paper and draw conclusions in \S \ref{conclusion}.


\section{Observations and Data Analyses}
\label{obs}

\begin{figure}[thb]  
\epsscale{0.9}
\plotone{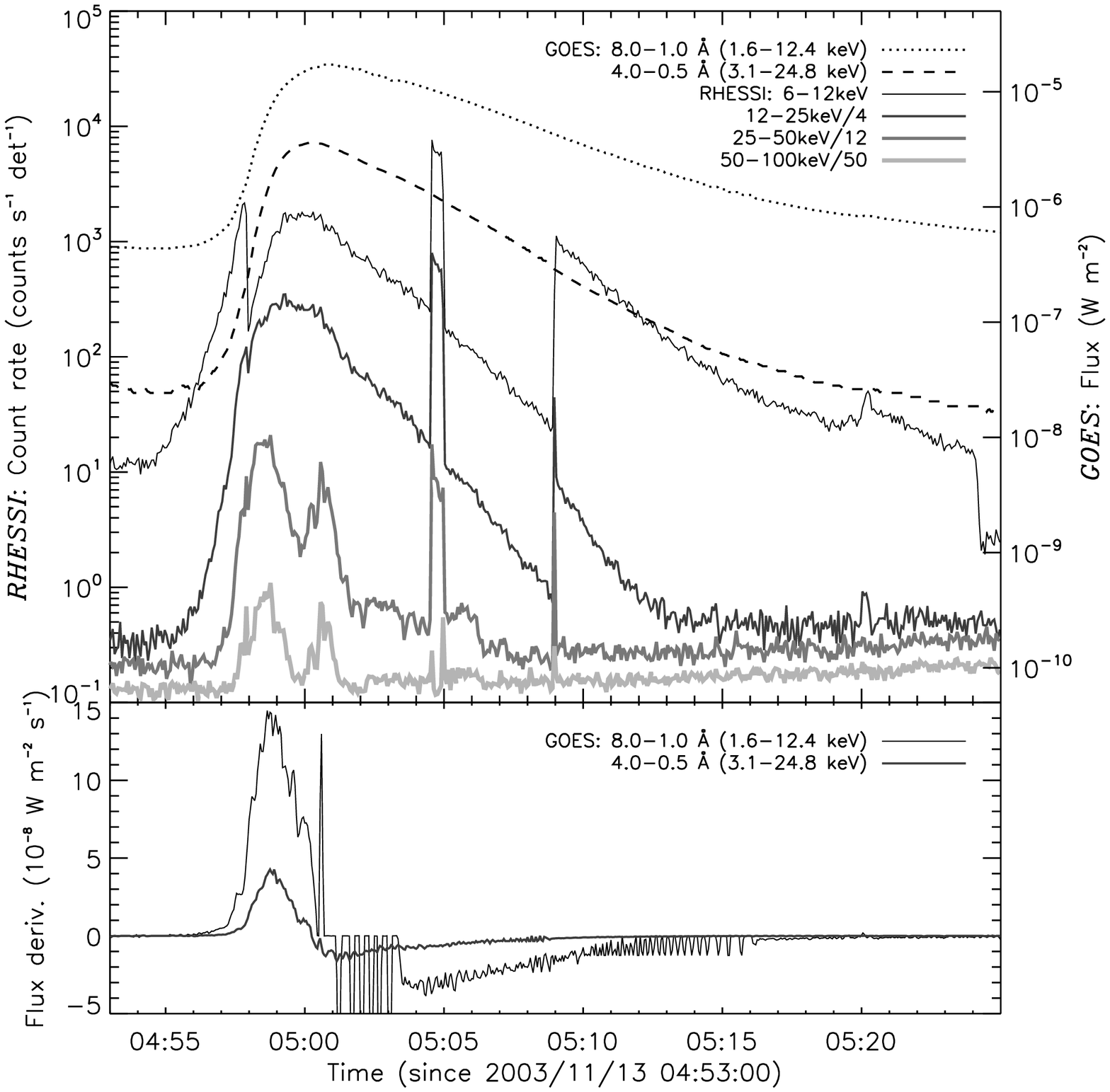}
\caption{{\it Top}: {\it RHESSI} and {\it GOES}-10 light curves. The {\it RHESSI} count rates 
are averaged over every 4 seconds, with scaling factors of 1, 1/4, 1/12, and 1/50 
for the energy bands 6-12, 12-25, 25-50, and 50-100 keV, respectively.
The sharp steps in the {\it RHESSI} light curves are due to
attenuator state changes, and the sudden drop of the 6-12 keV count rate 
near 05:24 UT results from the spacecraft eclipse.  The {\it GOES} fluxes in the bandpass of
8-1 {\AA} (1.6-12.4 keV) and 4.0-0.5 {\AA} (3.1-24.8 keV) are in a cadence of 3 seconds.
{\it Bottom}: Time derivative of the {\it GOES} fluxes.  Note that the periodic spikes of the low 
energy channel after 05:00:24 UT are calibration artifacts.}	
\label{lightcv}
\end{figure}

The flare under study is a {\it GOES} M1.7-class flare that occurred on 2003 November 13
in the Active Region 0501 after it appeared on the east limb. This event 
followed a period of extremely high solar activities in late October and 
early November when a series of X-class flares 
including the record setting X28 flare of 2003 November 4 took place \citep{Xu04, Liu04, Metc05, 
Vero06}. {\it RHESSI} had an excellent coverage of this flare.
Figure \ref{lightcv} shows the {\it RHESSI} and {\it GOES}-10 light curves. The {\it GOES} 
8-1 {\AA} (1.6-12.4 keV) and 4.0-0.5 {\AA} (3.1-24.8 keV) fluxes rise gradually and peak
at 05:00:51 and 05:00:15 UT, respectively. 
The {\it RHESSI} high energy ($>25$ keV) count rates, on the other hand, exhibit
two pulses peaking at 04:58:46 and 05:00:34 UT with the first one stronger. The steps in the 
{\it RHESSI} light curves are due to the attenuator (shutter) movements \citep{Lin02}. Before 
04:57:57 UT and after 05:08:59 UT, there were no attenuators in and between the two times the thin 
attenuator was in except for a short period near 05:05 UT 		
when the attenuator briefly moved out.

Figure \ref{mosaic} shows the evolution of the flare at different energies, which may
be divided into three phases: (1) before 04:57:57 UT, corresponding to the rising phase when the 
emission mainly comes from a flaring loop to the south. (2) Between 04:57:57 and 05:08:59 UT, 
the impulsive phase, 	
during which another loop to the north dominates the emission. This loop appears to 
share its southern footpoint (FP) with the loop to the south, 
which is barely visible because of its faintness 
as compared with the northern loop and {\it RHESSI}'s limited dynamic range of $\sim$10. 
(3) After 05:08:59 UT, the decay phase, when the shutters are out and two off-limb sources 
(identified as the LTs of the two loops) dominate. The relatively higher 
altitudes compared with earlier LT positions are 
consequences of the preceding magnetic reconnection, as seen in several other {\it RHESSI} 
flares \citep{Liu04, Sui04}. Clearly the southern loop, which extends to a relatively higher 
altitude, evolves slower and is less energetic than the northern one. We shall focus on 
the evolution of the northern loop during the first HXR pulse (04:58-05:00 UT) in this paper.


It is necessary to check if pulse pileup%
  \footnote{Two photons close in time are detected as one photon, having their energies added.
  Pileup of three or more photons is possible but at a much lower probability \citep{Smit02}.}
is important in this flare, 
before we can make more quantitative interpretation of the data.
The reason is that although we have applied the first order pileup correction \citep{Smit02}
in our spectral analysis, such a correction is challenging for images and not available 
at present. There are several ways to do the check, the detector livetime being
the first and simplest indicator. We first accumulated spatially integrated spectra 
for every 1 s time bins during the interval of 04:58:01-04:59:49~UT%
  \footnote{This time interval is also used in studying the evolution of the source morphology 
  in \S \ref{morp} (see text about Figure \ref{prof_3e}),
  which covers the bulk of the first HXR pulse.}, 	
using the front segments of all the nine detectors except detectors 2 and 7 
which have degraded energy resolution \citep{Smit02}.
We then obtained the livetime (between data gaps) from the spectrum object data
and averaged it over the seven detectors being used. The resulting livetime generally decreases 
with time, ranging from 96\% to 89\%, with a small modulation produced by the spacecraft spin. 
In this M1.7 flare, such a livetime is comparably high (cf. livetime of $\sim 55\%$ during 
the 2002 July 23 X4.8 flare and $\sim 94\%$ during the 2002 February 20 C7.5 flare)
and indicates minor pileup severity.

Another approach involves inspecting the change of the spectrum due to pileup.
We accumulated spectra over every 1 spacecraft spin period ($\sim 4$ s, with the same set
of detectors mentioned above) and used 
the pileup correction to obtain the relative fraction of the pileup counts among 
the total counts as a function of energy \citep{Smit02}. We find that the pileup counts 
amount to less than $\sim 10\%$ of the total counts at all the energies until 04:59:01~UT 
when the livetime drops to 91\%. After that, the relative importance of the pileup counts
continues increasing, but remains below $\sim 20\%$ of the total counts before 04:59:17~UT. 
Toward the end of the first HXR pulse
(04:59:45-04:59:49~UT, livetime $\sim 90\%$), the pileup counts to total counts
ratio exceeds $10\%$ in the entire 20-40~keV range and humps up to $43\%$ near 28 keV. 
We respectively integrate the pileup counts and total counts over the 20-40~keV band,
and plot their ratio versus time, as a general indicator of pileup severity 
(see Figure \ref{pileup}). Clearly this ratio is $\lesssim 15\%$ during the
first 2/3 of the interval shown and does not reach the moderate $\sim 25\%$ level
until the very end.

We therefore conclude that pileup effects are generally not very significant for this 
flare, especially during the first minute of the impulsive phase,
because the count rate is not too high and the thin shutter is in at times of interest
which further attenuates the count rate.
It should be noted that the two piled up photons (that result in a single photon seen in
the image) most probably originate from the same location on the Sun and pileup of 
photons across difference sources is relatively unimportant
(G. Hurford, private communication). Therefore the source geometry	
would not be significantly affected by pileup, except that there could be a ``ghost" 
of a low-energy source appearing in a high-energy image for very large (e.g. X-class) flares.
However, the spectra of individual sources derived from images are distorted,
which is relatively more significant at the LT than at the FPs. This is because 
ample low-energy photons dominate in population over high-energy photons,
having the highest probability to produce pileup, and generally most of the low-energy photons
are emitted by the LT source.


\subsection{Source Structure and Evolution}
\label{morp}

We now examine the images in greater detail. 
The {\it top left panel} of Figure \ref{mdi} shows {\it RHESSI} 
CLEAN \citep{Hurf02} images of the northern loop at 
9-12, 12-18, and 28-43 keV for 04:58:22-04:58:26 UT. (Although the 4-second integration time 
is rather short, the image quality is reliable with a well-defined source structure.) At 9-12 keV the 
LT dominates and the emission extends towards the two FPs, which dominate the emission at 28-43 
keV and above with the northern FP (N-FP) much brighter than the southern one (S-FP). One of 
the most interesting features of the source structure is that 
emission from the legs of the loop dominates at the intermediate energy (12-18 keV). 
Similar structures are also observed for several other 
time intervals during the first HXR pulse (see discussions below). 
We find emission from the legs is a transient phenomenon at intermediate energies, 
because when integrating over a long period and/or a broad energy band, 
the LT and/or FP sources become dominant. To our 
knowledge no images like this have been reported before. We attribute this in part to the 
relatively short integration time and {\it RHESSI}'s high energy resolution.

To be compared with observations at other wavelengths, 
the same images at 9-12 and 28-43 keV (solid contours) are shown with the {\it SoHO}/EIT,
MDI magnetogram, and MDI white light maps in the other three panels of Figure \ref{mdi},
where the dashed contours depict the southern loop at 6-9 keV for 04:57:40-04:57:52 UT. 
The EIT image at 04:59:01 UT ({\it upper right panel}) shows emission at 195 {\AA} co-spatial
with the SXR emission from the northern loop. 
The brightest 195 {\AA} emission, an indicator of the highest differential emission measure 
(thus the highest density) at $\sim 1.3 \times 10^6$ K,
appears to be close to the N-FP which is also the strongest FP in HXRs%
  \footnote{EIT 195 {\AA} passband images have a relatively narrow temperature response range 
  with a characteristic temperature of $1.3\times 10^6$ K \citep[see][Fig. 12]{Dere00},
  and emission intensity would be lower for both higher and lower temperatures.}.  
The {\it lower left panel} displays the X-ray emission along with the post-flare (05:57 UT) 
MDI magnetogram. This clearly shows that the northern loop straddles across
a polarity reversal with the brighter N-FP associated with a stronger magnetic field%
 \footnote{Note that since this flare occurred near the solar limb, 
  the line-of-sight magnetogram measures mainly the horizontal (parallel to the solar surface)
  component of the magnetic field. The vertical component is more relevant here because
  flaring loops are usually perpendicular to the surface. 
  However, it would be reasonable to assume that the vertical component scales
  with the horizontal one, and the polarity reversal line in the latitudinal direction is
  essentially not subject to the line-of-sight projection effect, as it seems very likely here.}.
The southern loop (dashed contours) is associated with a even weaker magnetic field.
Here we show the MDI magnetogram recorded one hour after the flare's impulsive phase because
during a flare there are many uncertainties in the magnetic field measurement.
The {\it lower right panel} shows the MDI continuum map at 12:47 UT 
(about 8 hours after the flare), suggesting that the
flare occurred above the lower sunspot region (dark area).
Note that during this interval the sunspot has moved westward for about $4^{\circ}$ 
in heliographic longitude.	
We do not plot the MDI white light map at the time of the flare because then the sunspot was 
nearly on the limb and barely visible.

\begin{figure}  
\epsscale{0.93}  
\plotone{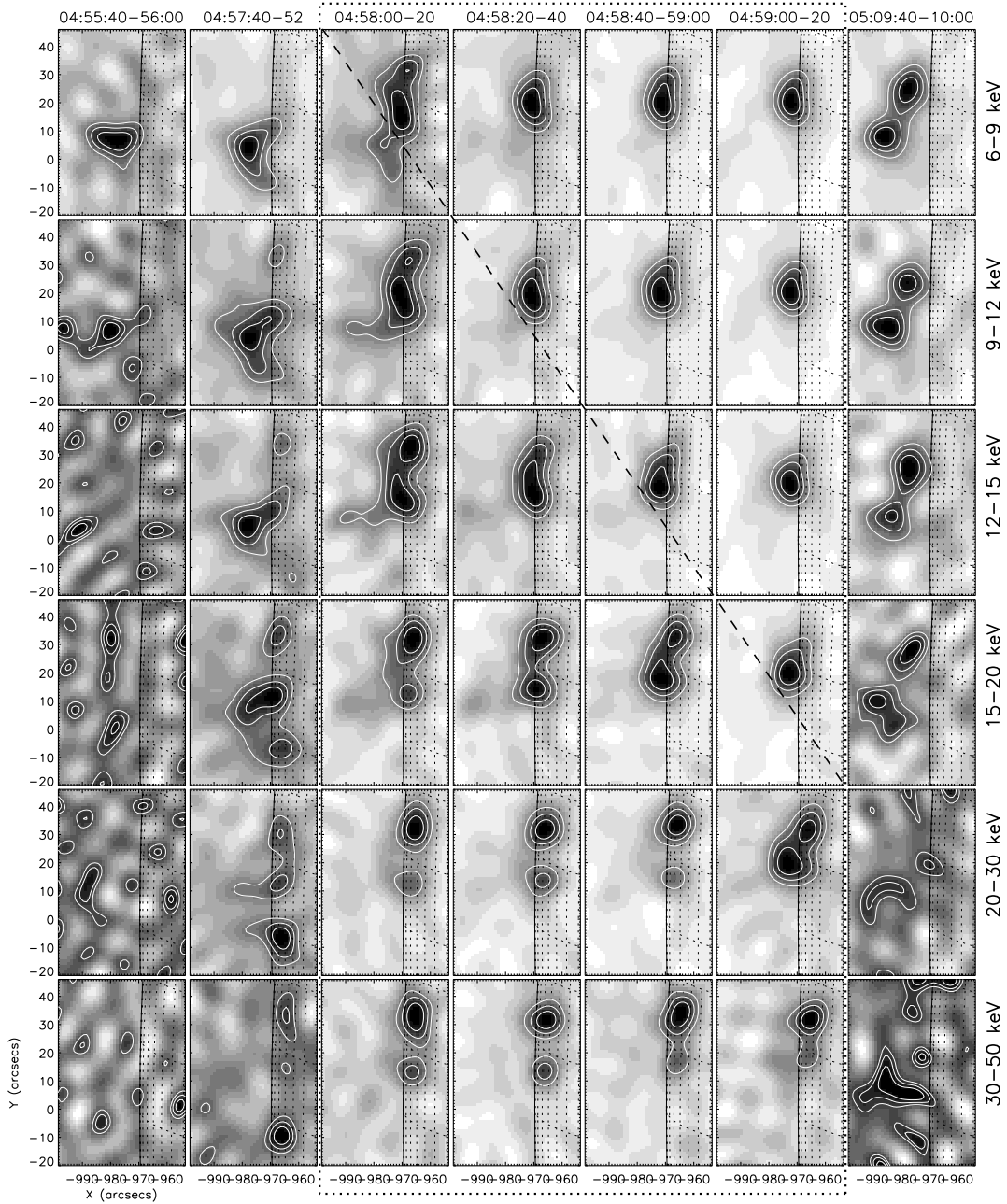}
\caption{Mosaic of CLEAN images at different energies (rows) and times (columns). 
Contour levels are set at 40, 60, and 80\% of the maximum brightness of each image.
The front segments of detectors 3-6 and 8 were used for reconstructing these images
and the others presented in this paper, yielding a spatial resolution of $\sim$$7\arcsec$.
We selected the integration intervals to avoid the times when the attenuator state changed.
The large dotted box encloses the images during the first pulse of the impulsive phase,
and within this time interval the dashed diagonal line separates the frames showing double sources 
or an extended source from those with a compact single LT source. 
}
\label{mosaic}
\end{figure}

\begin{figure}[thb]  
\epsscale{0.8}
\plotone{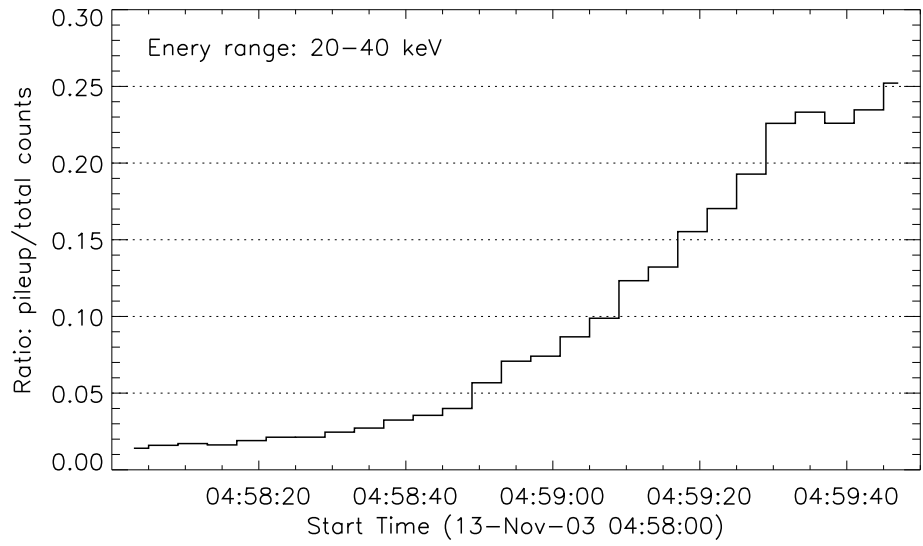}
\caption{The ratio of pileup counts to total counts, both integrated over the 20-40 keV range
in time bins of 1 spacecraft spin.}
\label{pileup}
\end{figure}

\begin{figure}[thb]  
\epsscale{0.45}		
\plotone{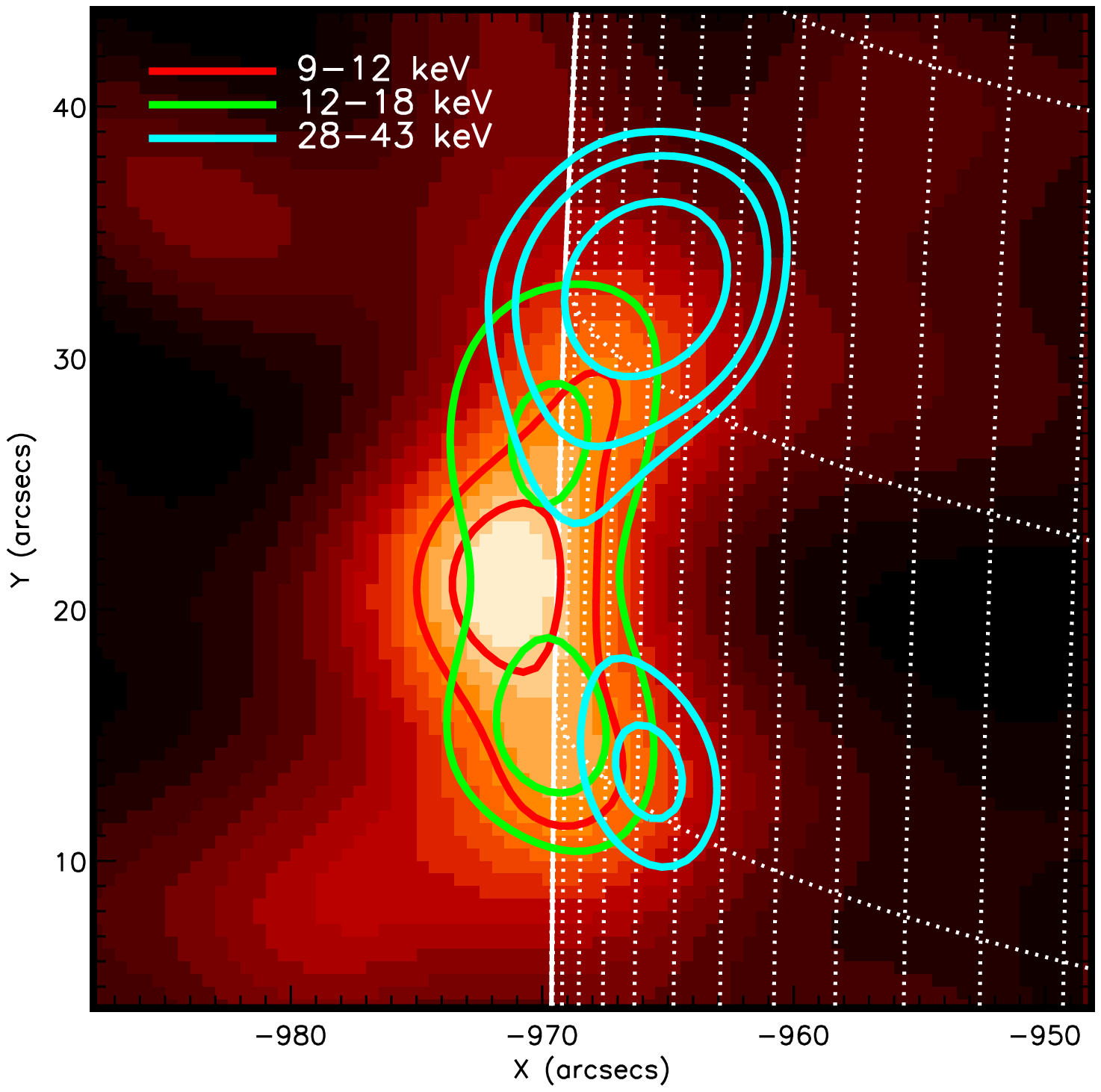}	
\plotone{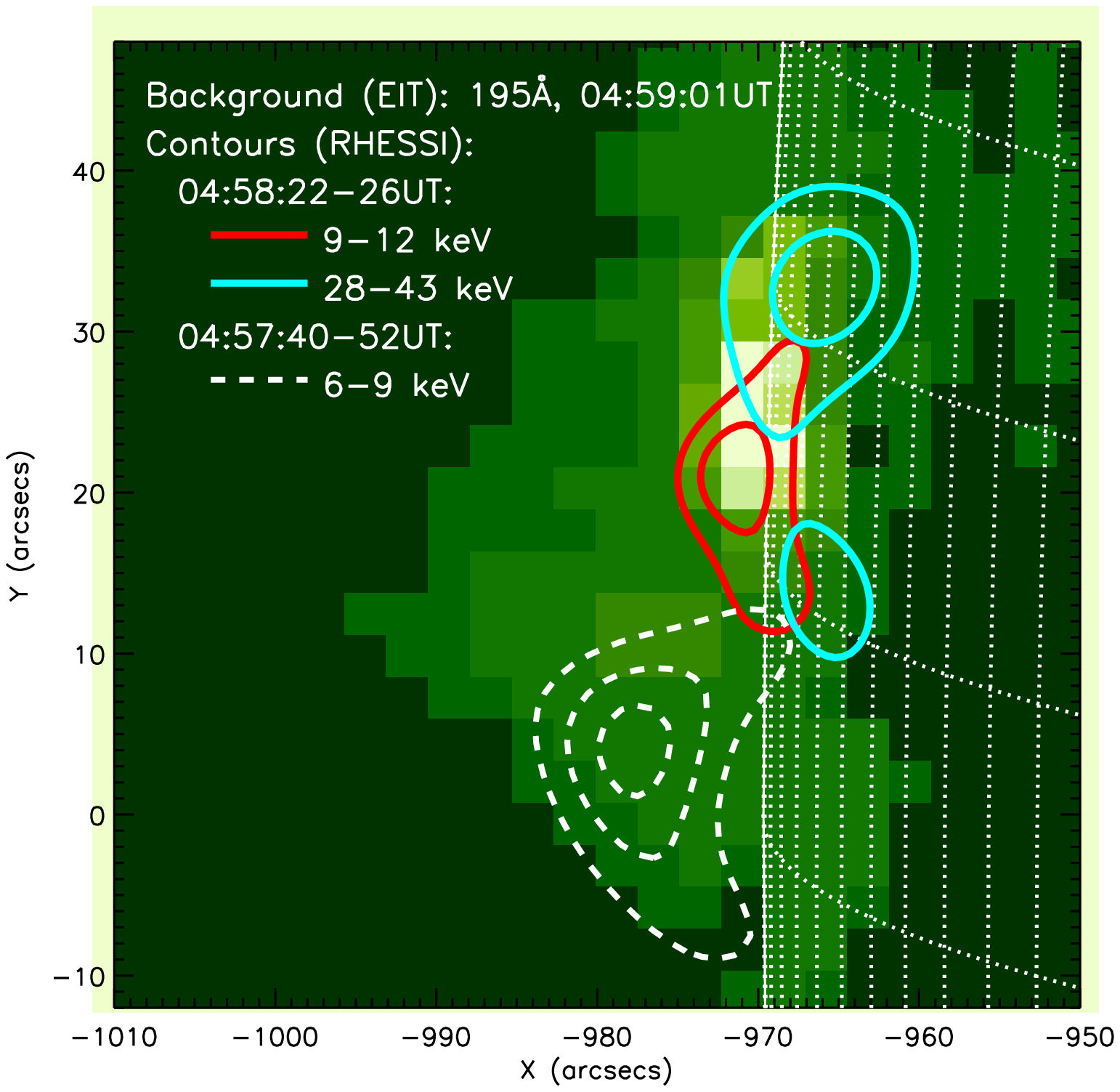}	
\plotone{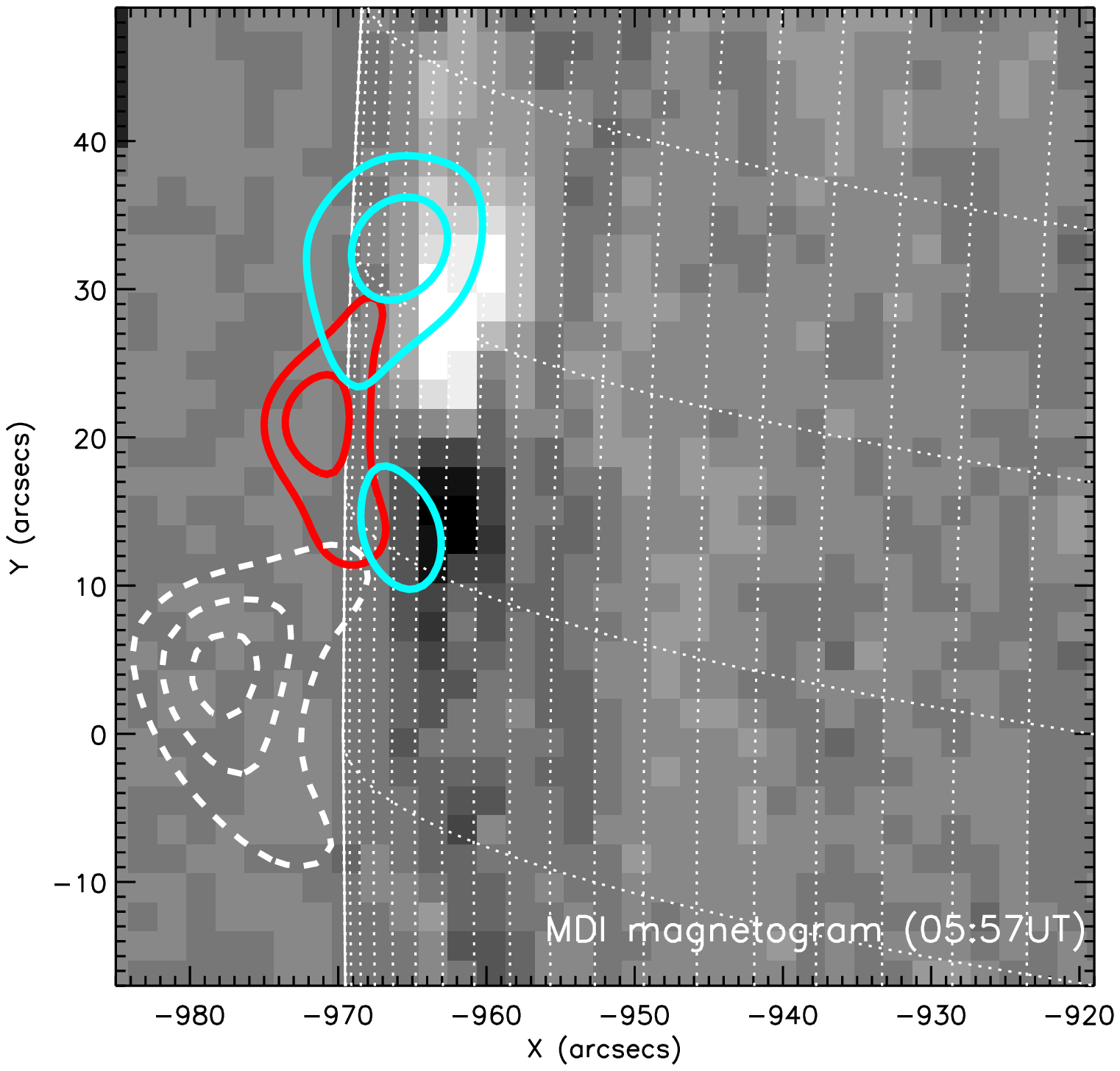}
\plotone{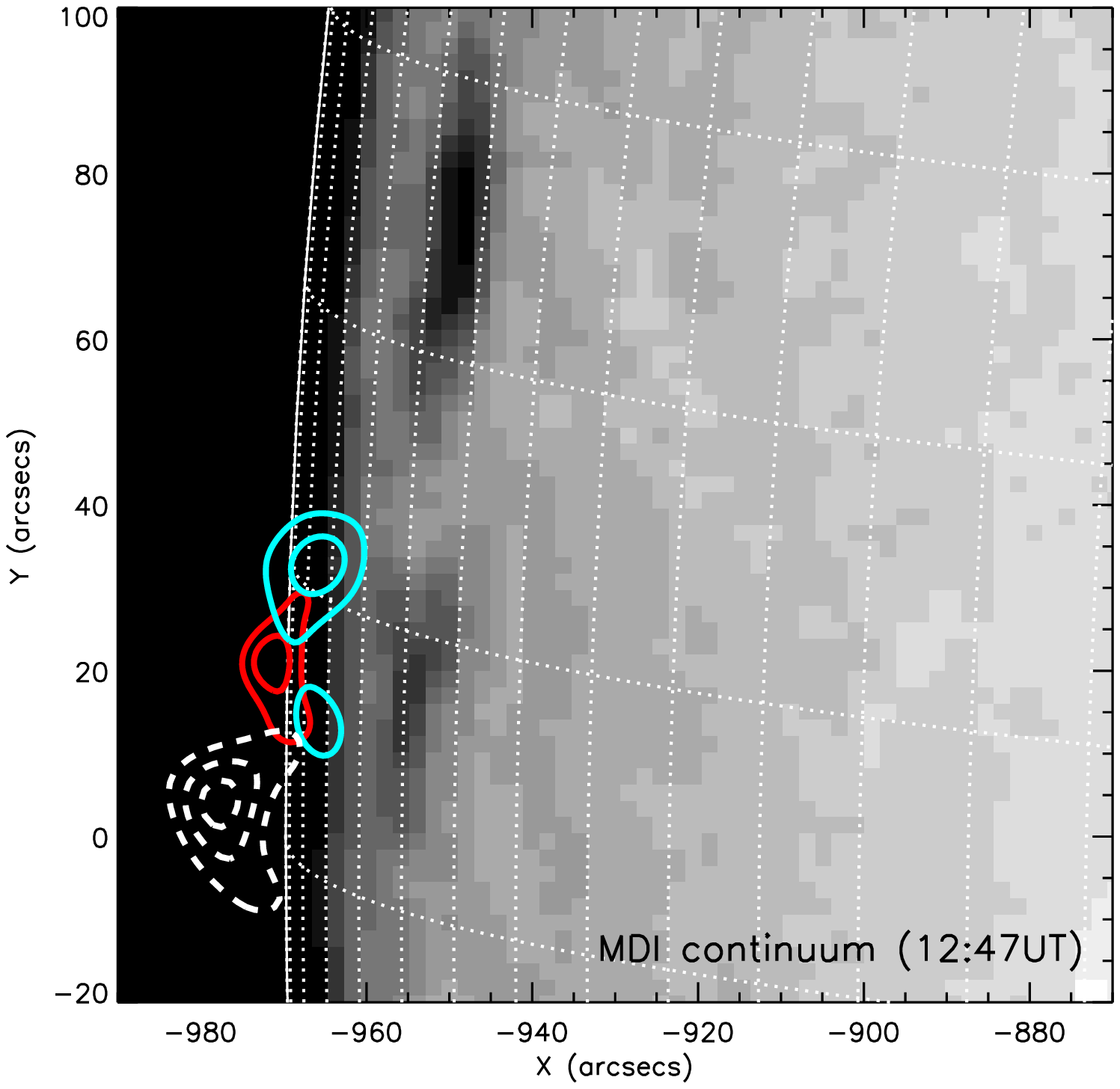}
\caption{{\it Top left}: {\it RHESSI} images for 04:58:22-04:58:26 UT during the first HXR pulse.  The 
background is the image at 9-12 keV. The contour levels are at 75 and 90\% for 9-12 
keV, 70 and 90\% for 12-18 keV, and 50, 60, \& 80\% for 28-43 keV. 
  {\it Top right}: an EIT 195 {\AA} image at 04:59:01 UT, 
showing co-spatial EUV emission in the northern HXR loop. 
The solid contours are the same as in the {\it top left panel} at 9-12 and 28-43 keV except that 
the contour levels are 50 and 80\% for the latter.  A 6-9 keV {\it RHESSI} image 
(same as the second panel in the first row of Figure \ref{mosaic}) for 04:57:40-04:57:52 
UT is plotted as dashed contours (at 50, 70, 90\% levels) which depict the southern loop.
The same set of contours is plotted in the two bottom panels as well.
  {\it Bottom left}: an MDI magnetogram at 05:57 UT. The line-of-sight magnetic field in the map ranges from 
-351 G (black: away from the observer) to 455 G (white) with the FPs near the strong magnetic 
field regions. 
  {\it Bottom right}: an MDI continuum map at 12:47 UT showing the sunspots. The heliographic grid spacing
is $2^{\circ}$.}
\label{mdi}
\end{figure}

Next we consider the evolution of the northern loop. We notice that, 
as shown in the four columns for 04:58:00-04:59:20 UT (boxed by the dotted line)
in Figure \ref{mosaic}, the FPs initially appear at all energies but later on
dominate only in the high energy bands, while the LT is first evident at low energies 
and becomes more and more prominent at relatively higher energies as indicated by the dashed diagonal 
line. The emission from the LT also extends towards the legs at intermediate energies and in a 
given energy band the emission concentrates more and more at the LT with time. 
These are expected to be common features of flares 
with a single loop because of chromospheric evaporation that can
increase the plasma density in the loop making the LT dominate at progressively higher energies. 
However, because the 20 second integration time is relatively long, these images do not uncover 
the details of the evaporation process. To remedy this we have carried out three different but
complementary analyses of the images with higher temporal or energy resolution.

\begin{figure}[thb]  
\epsscale{1.0}	
\plotone{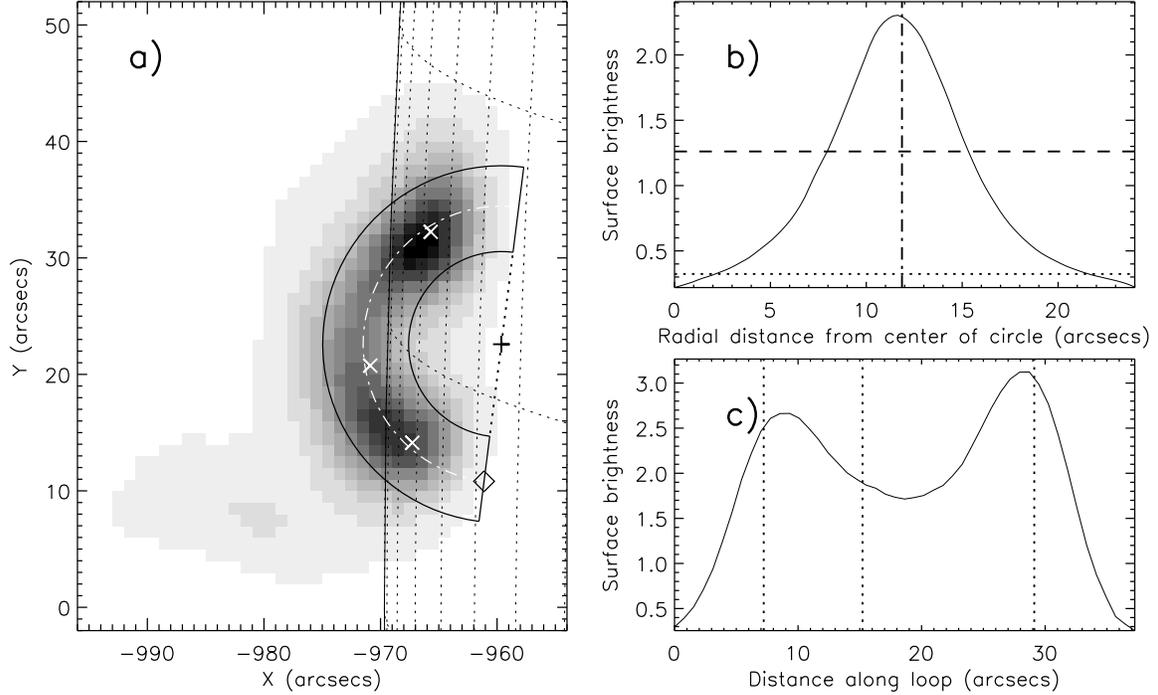} 	
\caption{({\it a}) Synthesized image obtained by superimposing 30 8-second images between 
04:58:08 and 04:58:56 UT for 5 energy bands: 9-12, 12-15, 15-20, 20-30, and 30-50 keV. 
The three crosses mark the LT and two FPs identified as the emission centroids of the 
corresponding sources in the 04:58:12-04:58:53 UT images at 6-9 keV and 50-100 keV, 
respectively. The solid lines represent 
the semi-circular model loop with the center of the circles marked by the plus symbol. 
The white dot-dashed line is the central arc (see below) of this loop  
and the diamond indicates the start point of the distance in {\it panel c}. 
  ({\it b}) Radial brightness profile averaged along the loop, obtained from the
image shown in {\it panel a}. The distance is measured from the 
center of the circles.  The horizontal dashed line marks the 50\% level of the 
maximum and the crossings of this line with the profile define the radii of 
the two solid semi-circles	
in {\it panel a}.  The 5\% level is represented by the 
horizontal dotted line. The vertical dot-dashed line denotes the radial position
of the central arc of the loop.  
  ({\it c}) Same as {\it panel b}, but for surface brightness along the 
loop's central arc, averaged perpendicular to the loop. 
The three vertical dotted lines mark the corresponding positions of the cross signs in {\it panel a}.
}
\label{loop-model}
\end{figure}

1. To study the source morphology change over short time intervals, we model the loop geometry and study the 
evolution of the HXR brightness profile along the loop. We first made CLEAN images
in two energy bands of 6-9%
  \footnote{Since the thin attenuator was in at that time, counts below 10 keV are 
  likely dominated by photons whose real energy is about 10 keV higher than the detected
  energy. This is due to strong absorption of lower energy ($< 10$ keV) photons by the attenuator 
  and escape of the germanium K-shell fluorescence photons that are produced by photoelectrical 
  absorption of higher energy (10-20 keV) photons in the germanium detector \citep[see][\S 5.2]{Smit02}. 
  However, for the flare under study, the 6-9 keV image most likely reveals the real LT
  morphology, because there are ample thermal photons at lower energies originating from the LT source
  and photons at slightly higher energies seem to come from the same location.}
and 50-100 keV over the time interval of 04:58:12-04:58:53 UT 
that covers the plateau portion of the first HXR pulse. From these two images 
we obtained the centroids (indicated by the white crosses in Figure \ref{loop-model}{\it a}) 
of the sources identified as the LT (6-9 keV) and two FPs (50-100 keV), respectively. 
Assuming a semi-circular loop that connects the three centroids,
we located the center of the circle marked by the plus sign in Figure \ref{loop-model}{\it a}. 
The grey scale in Figure \ref{loop-model}{\it a} was obtained by 
superposition%
 \footnote{Because we are interested in determining the average loop geometry during the first pulse 	
 when the low energy X-ray flux has changed dramatically, using this approach to map the loop will 
 ensure a relatively uniform brightness profile along the whole loop by assigning equal weights
 to images at different energies.
 On the other hand, if one simply integrates over the entire time range of 04:58:08-04:58:56 UT and 
 energy band of 9-50 keV, the source morphology will be dominated by the LT source that emits most 
 of the photons at a later time and relatively lower energies, which may not properly depict the 
 loop geometry during the HXR pulse.}
of 30 images (six 8-second intervals from 04:58:08 to 04:58:56 UT in five energy bands:  
 9-12, 12-15, 15-20, 20-30, and 30-50 keV) reconstructed with the 
PIXON algorithm \citep{Metc96, Hurf02}.
Figures \ref{loop-model}{\it b} and \ref{loop-model}{\it c} respectively show the intensity profiles
perpendicular to and along the loop (averaged over the respective orthogonal directions).
The inner and outer circles (at $r=8\farcs0$ and $15\farcs3$) in Figure \ref{loop-model}{\it a} 
show the positions of the 50\% of the maximum intensity in Figure \ref{loop-model}{\it b}.
However, for inferring the intensity profile along the loop we use radially integrated flux
down to the 5\% level.
This enables us to include as much source flux as possible (with little contamination from the 
southern loop). We define the mean of the radii at the 5\% level as the radius of the
central arc of the loop (the white dot-dashed line in Figure \ref{loop-model}{\it a}).

\begin{figure}[thb]  
    \epsscale{0.32}	
    \plotone{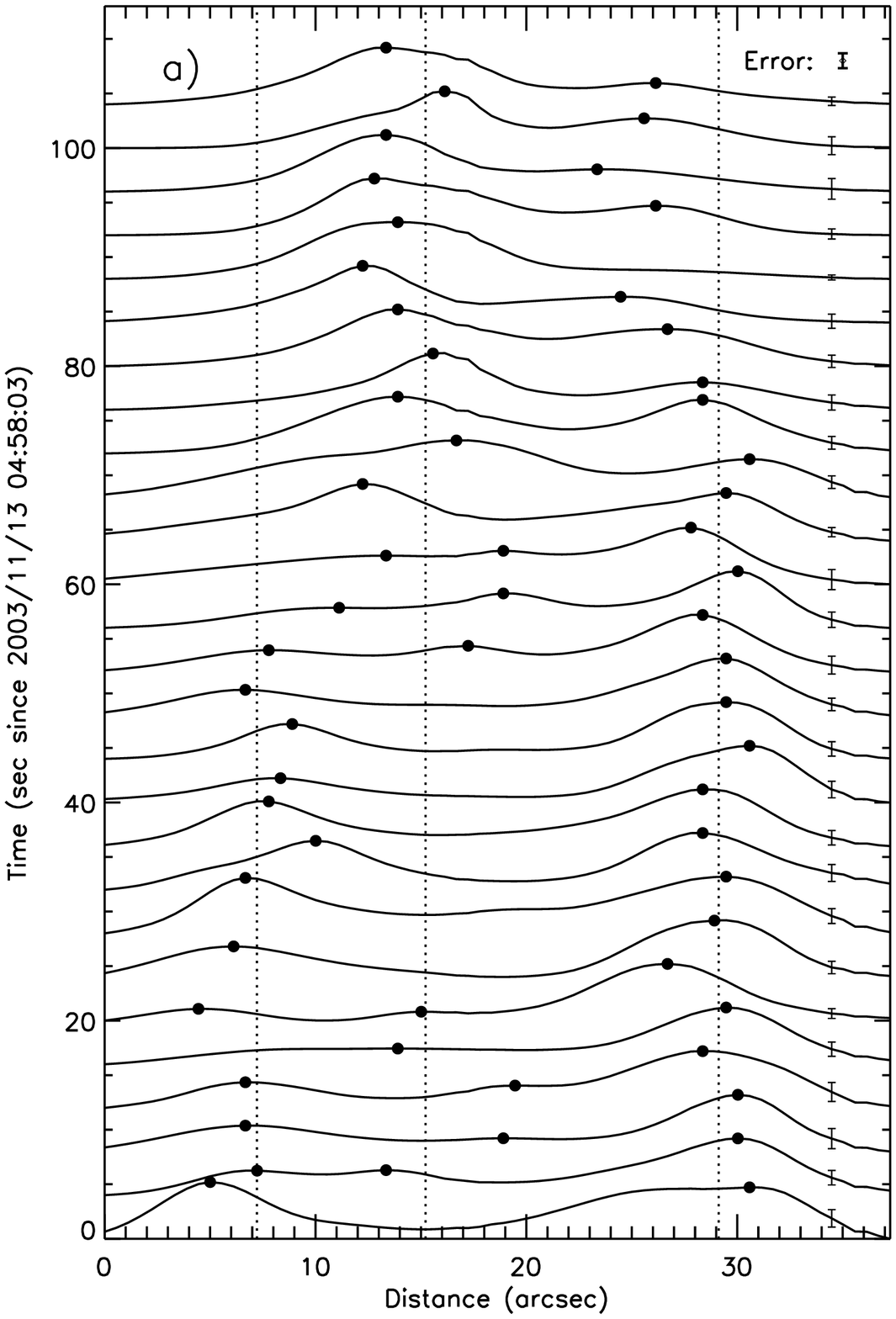}
    \plotone{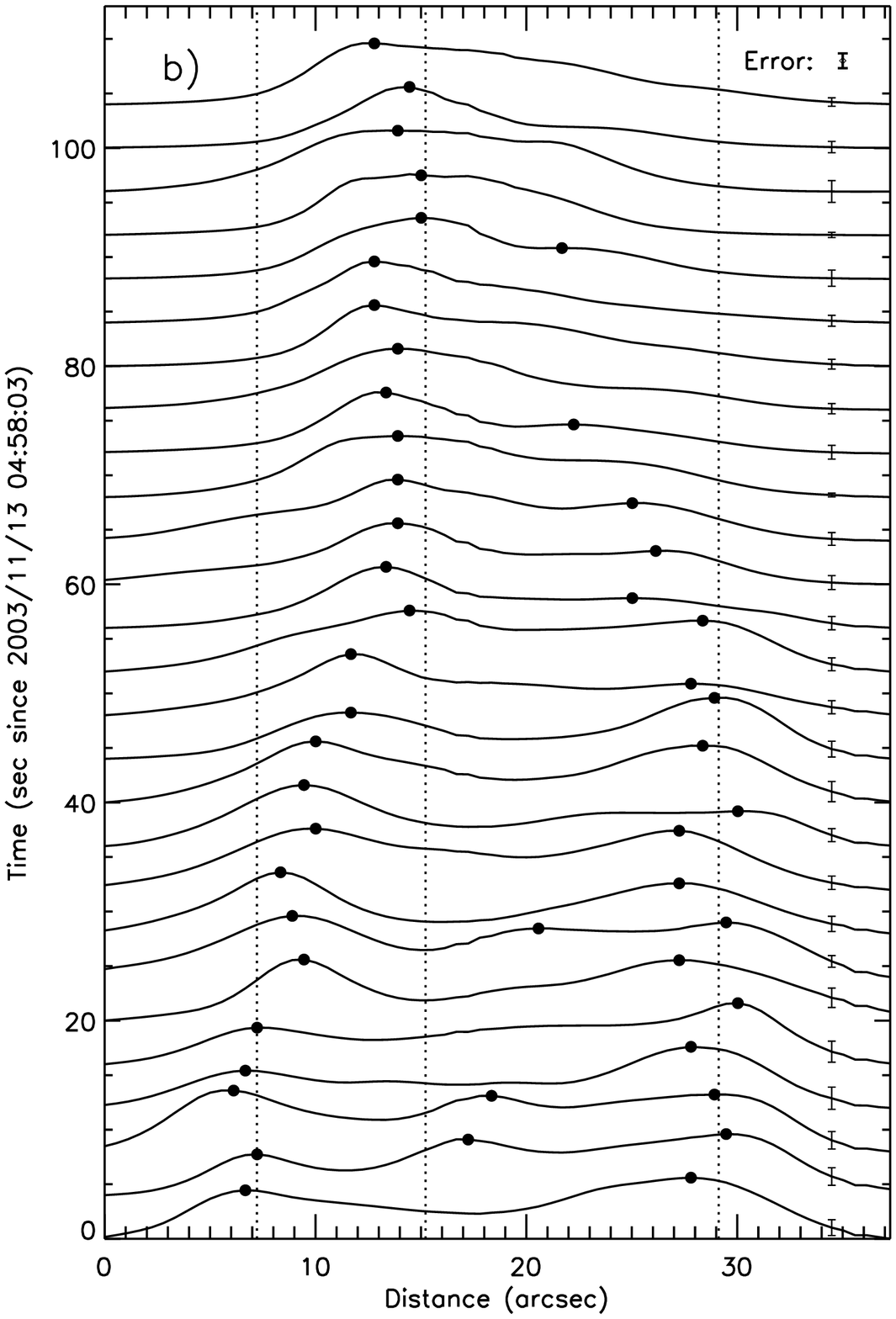}
    \plotone{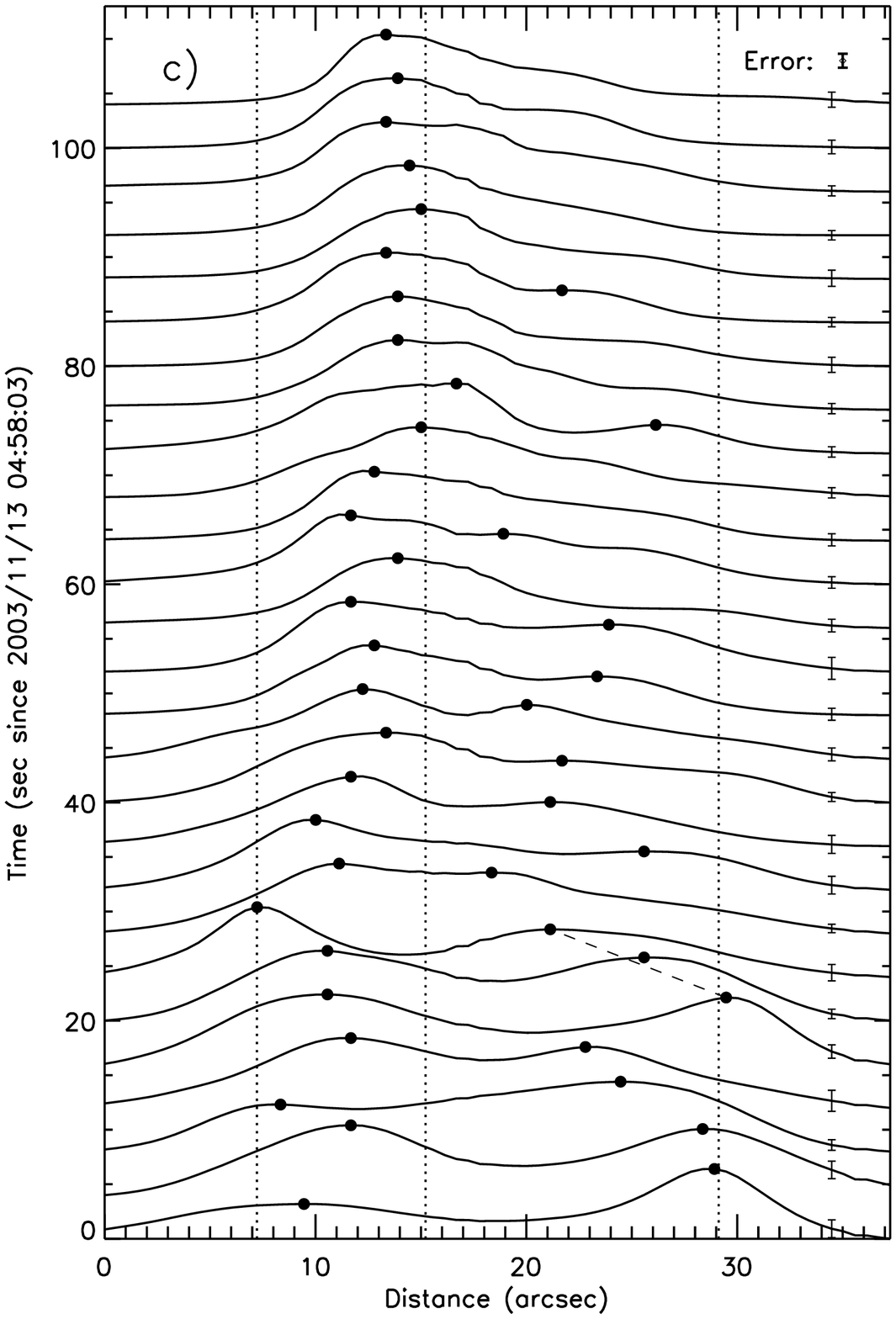}
    \caption{
    ({\it a}) {\it left}: Evolution of the 20-30 keV brightness profile along the loop in a cadence of 
    4 seconds starting at 04:58:03 UT. Each profile is normalized to its own maximum and has an 
    integration time of 1 spacecraft spin period ($\sim$4 seconds) whose central time is used to 
    label the vertical axis. The filled circles mark the local maxima and the three vertical 
    lines are the same as those in Figure \ref{loop-model}{\it c}.
    The error bar on each curve indicates an estimated uncertainty of the profile and
    the stand-alone error bar in the upper-right corner represents the overall uncertainty
    (13\%) of all the profiles.
    ({\it b}) {\it middle} and ({\it c}) {\it right}: Same as {\it panel a} but for 15-20 
    and 12-15 keV, with an overall uncertainty of 12\% and 10\%, respectively. 
    With the dashed straight line in {\it panel c}, 
    we estimate the speed of the emission maximum at $\sim$$10^3$ km s$^{-1}$. 
    Note the slightly different scales among the three panels
    for the profiles and their error bars.
    }
    \label{prof_3e}		
  \end{figure}

With the above procedure, one can study the evolution of the brightness profile along 
the loop at different energies. Figure \ref{prof_3e}	
shows the results obtained from PIXON images with 	
an integration time of 1 spacecraft spin period ($\sim$4 seconds) 	
from 04:58:01 to 04:59:49 UT for three energy bands (20-30, 15-20, and 12-15 keV). 
Using a simple algorithm we determine the local maxima
whose slopes on both sides exceed some threshold value and mark them by filled circles. 
We compare each profile with its counterpart obtained from the CLEAN image
(with the same imaging parameters), and use the RMS of their difference to estimate 
the uncertainty as indicated by the error bar near the right-hand end of the 
corresponding profile. For each panel, the RMS difference of all the profiles, 
as a measure of the overall uncertainty, is shown as the error bar in the upper-right 
corner. This uncertainty is about 10\% for the three energy bands;
as expected, it increases slightly at higher energies which have lower counts.

Figure \ref{prof_3e}{\it a}		
displays the profile at 20-30~keV, which as expected (see Figure \ref{mosaic})
shows emission from the two FPs with fairly constant positions until the very last stage
when the LT emission becomes dominant%
  \footnote{As noted earlier, pulse pileup in the 20-40~keV range becomes relatively important 
  at this very late stage, which means that a fraction of the 20-30~keV photons seen
  in the image are actually piled up photons at lower energies.
  }.
At this stage, the S-FP becomes undetectable and the N-FP has moved very close to the LT. 
At the lower energy (15-20 keV, Figure \ref{prof_3e}{\it b})		
the maxima tend to drift toward the LT gradually and eventually merge
into a single LT source. At the even lower energy (Figure \ref{prof_3e}{\it c})		
this trend becomes even more pronounced and the drift starts earlier except here the shift is
not monotonic and there seems to be a lot of fluctuations. 
We also repeated the same analysis at a higher cadence (every 1 second, $\sim 4$ second
integration interval) with both PIXON and CLEAN algorithms. The evolution of the 
resulting profiles (although oversampled and thus not independent for neighboring profiles)
appears to be in line with that shown here at a 4-second cadence obtained with PIXON.
The general trends of
these results indicate that high energy HXR producing electrons lose their energy
and emit bremsstrahlung photons higher and higher up in the loop as the flare progresses. 
This can come about simply by a gradual increase of the density in the loop, 
presumably due to evaporation of chromospheric plasma. From the general drift of
the maxima we obtain a time scale ($\sim$10's of seconds) and a velocity of
a few hundred km s$^{-1}$ consistent with the sound speed or speed of slow magnetosonic
waves. As stated above
at low energies we see some deviations from the general trend, some of which do not appear
to be random fluctuations. If so and if we take one of the evident shorter time
scale trends shown by the dashed line in Figure \ref{prof_3e}{\it c}	
we obtain a large velocity%
\footnote{cf. Among the highest observed up-flow velocities in chromospheric evaporation are about 
$10^3$ km s$^{-1}$ \citep{Anto90} and 800 km s$^{-1}$ \citep{Dosc94}, obtained from
blue-shifted Fe XXV spectra. }
 ($\sim$$10^3$ km s$^{-1}$) which is comparable to the Alfv\'{e}n or fast magnetosonic
wave speed. This may indicate that another outcome of energy deposition by
non-thermal particles is the excitation of such modes which then propagate from the
FPs to the LT
  and might be responsible for the circularly polarized zebra pattern
  observed in the radio band \citep{Cher05}.   
This, however, is highly speculative because the spatial resolution ($\sim$$7\arcsec$)
is not sufficiently high for us to trust the shorter time scale variation.
The longer time scale general trend, however, is a fairly robust result.

\begin{figure}[thb] 
\epsscale{0.7}
\plotone{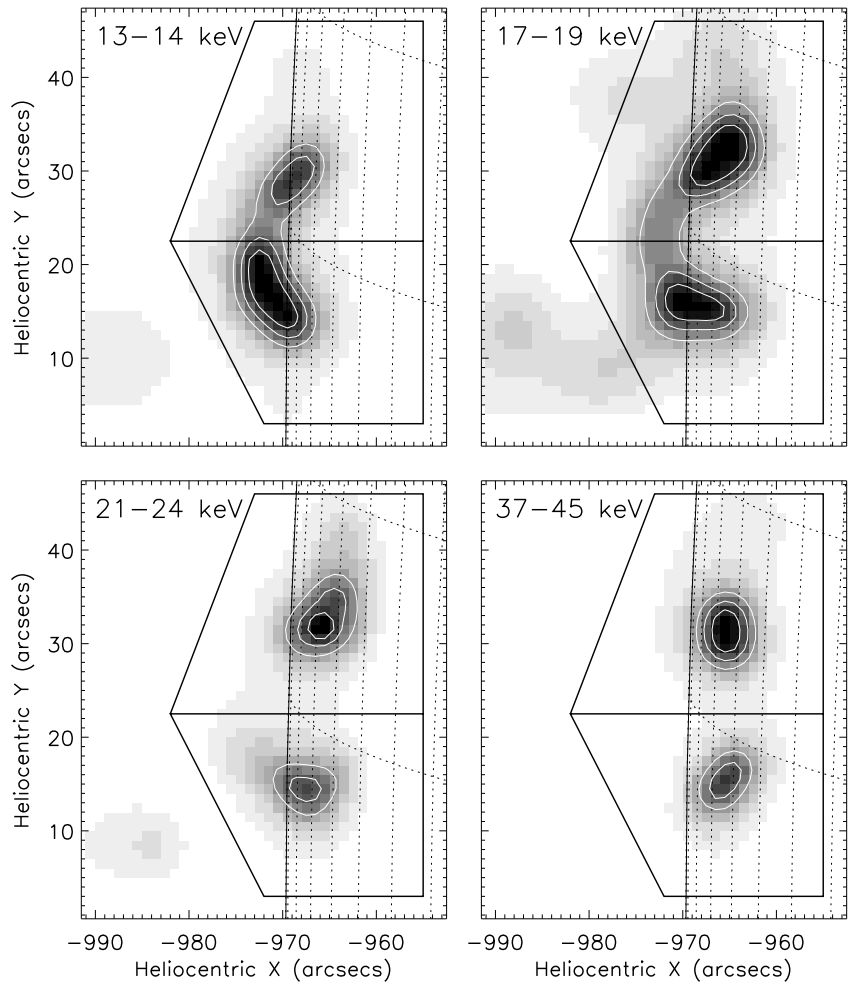}
\caption{PIXON images at 04:58:24-04:58:48 UT in different energy bands. The overlaid 
boxes were used to divide the loop into halves to calculate the corresponding centroids.}
\label{box}
\end{figure}

2. Instead of examining the source structure with high time resolution, we can investigate
it with higher energy resolution at longer integration intervals 
as tradeoff for good count statistics and image quality.
To this end, we have made PIXON images during three consecutive 24 s intervals 
starting from 04:58:00 in 20 energy bins within the 6-100 keV range.
Figure \ref{box} shows a sample of these images at 04:58:24-04:58:48 UT.
Figures \ref{prof_24s}{\it a}-\ref{prof_24s}{\it c} show the X-ray emission profile along the loop at 
different energies for the three intervals%
  \footnote{Note that pileup effects, as discussed earlier, are insignificant 
  during this period of time (see Figure \ref{pileup}).}. 
As in Figures \ref{prof_3e}{\it a}-\ref{prof_3e}{\it c}		
the high energy emission is dominated by the FPs but there is a decrease of 
the separation of the FPs with decreasing energies and with time. Again at later stages
the LT dominates and the profile becomes a single hump. 
The general trend again suggests an increase of the gas density in the loop.
At lower energies ($< 15$ keV), the profile is more complicated 
presumably due to many physical processes (in addition to chromospheric evaporation), such as 
thermal conduction and transport of high energy particles, thermal and non-thermal bremsstrahlung,
wave excitation and propagation, wave-particle coupling, and even particle acceleration, 
which may be involved. 
We believe that a unified treatment of acceleration and HD processes with 
physical conditions close to the flare is required for interpreting 
these results to uncover the details. 

To quantify this aspect of the source structure evolution, we divided the loop 
into two halves as shown by the boxes in Figure \ref{box} and calculated their emission centroids.  The 
resulting centroids at the three times together with the central arc of the model loop are plotted in 
Figure \ref{centroids}.  As can be seen, for each time interval the centroids are distributed along 
the loop with those at higher energies being further away from the LT, and the entire pattern shifts 
toward the LT with time. Figure \ref{centr-dist} shows the centroid positions of the northern 
half of the loop (where the source motions are more evident) along and perpendicular to the loop 
during the three intervals. This again shows that higher energy emission is farther 
away from the LT and the centroids shifted towards the LT with time,
but similarly there are some complicated patterns at low and intermediated energies. 
All these are consistent with the general picture 
proposed above for the chromospheric evaporation process.

3. To further quantify the source motions, we obtained the brightness-weighted 
standard deviation or the 2nd moment of the profiles.
In general the moment measures the compactness of the overall emission, 
but does not yield the sizes of individual sources whose measurement is
still challenging for {\it RHESSI} \citep{Schm02}. Hence our attention should be
paid to the general trend of the moment rather than its absolute values, which
may be subject to large uncertainties and thus less meaningful.
The moments of the profiles resulting from CLEAN images (in three energy bands over 8 s intervals)
are plotted in Figure \ref{time}{\it b}.	
There is a general decrease of the moment 
with the decline starting earlier at lower energies. Such a decrease is expected
if the two FPs move closer to each other. However, caution is required here 
because a decrease of this quantity could also come about by other causes, say by
an increasing dominance of the brightest source.
We therefore checked the original images and the corresponding profiles when 
interpreting our results. 
To estimate the uncertainty of the moment, for each energy band we repeated the calculation with 
different integration time ({\it panel c}).  The resulting moments remain essentially unchanged
and as excepted the fluctuations of the moment decrease with increasing integration time. 
We also plot in {\it panel c} the moment (solid line) obtained from PIXON images 
with an integration time interval of two spin-period ($\sim 8$ s), 
which basically agrees with its CLEAN counterpart in the general trend. The gradual%
  \footnote{On the other hand the jumps (if real) of the moment may suggest a transient phenomenon.}
decrease of the moment is consistent with the motion of the centroids of sources
up along the legs of the loop, which can take place by a continuous increase of the 
gas density in the loop due to evaporation.


\subsection{Spectral Analysis}
\label{spec}

Spectral analysis can be used to study the evaporation process as well. With an isothermal plus a 
power law model, we fitted the spatially integrated {\it RHESSI} spectra down to 6 keV 
\citep{Smit02} for every 8 second interval during the impulsive phase. The emission measure ($EM$) and 
temperature of the isothermal component (asterisk symbols) are plotted in {\it panels d} and {\it e} 
of Figure \ref{time}, respectively. The $EM$ rises almost monotonically with 
time from 0.6 to 14.2 $\times 10^{49}$ cm$^{-3}$.  	
This translates into an increase of the plasma density ($n=\sqrt{EM/V}$) by a factor of $\sim$5
assuming a constant volume $V$. The temperature remains almost constant with a trend of slight 
decrease with time. The $EM$ and temperature derived from the {\it GOES} data (plus symbols) are 
also shown for comparison. In general the {\it GOES} results are smoother and the temperature 
increases monotonically but remains below that of the {\it RHESSI}, consistent with previous 
results \citep{Holm03}. This is expected because {\it RHESSI} is more sensitive to 
higher temperatures than {\it GOES}. However, surprisingly, the {\it GOES} emission measure is also lower 
than that of {\it RHESSI} as opposed to what is the case more generally 
(see Holman et al. 2003). It is not clear whether or not
this is due to a problem related to the {\it RHESSI} calibration at low energies. 
Nevertheless, the continuous increase of the $EM$'s at comparable rates does suggest a 
gradual increase of the plasma density.

The best fit parameters of the power law component with a low energy cutoff
are plotted in Figure \ref{time}{\it f}. 
The power-law index $\gamma$ (plus symbols) is anti-correlated with the high energy light curves 
(see Figure \ref{time}{\it a}) and shows a soft-hard-soft behavior.  It starts with 4.43 at
04:58:02 UT, drops to 3.82 at the impulsive peak (04:58:26 UT), and rises up to 7.12 at
04:59:46 UT. The high indexes ($> 5$) may be an indicator of high-temperature thermal
rather than non-thermal emission. Thus in what follows we limit our analysis to time up to 04:59:20 UT.
The low-energy cutoff (asterisks) of the power law is about $15$ keV
and is near the intersection of the isothermal (exponential) and power-law components.



\subsection{The Neupert Effect}
\label{Neupert_section}

The Neupert effect is commonly quoted as a manifestation of chromospheric evaporation 
\citep{Denn93} and a simple energy argument (e.g. Li et al. 1993) is often used to account for 
the relationship between SXR and HXR fluxes ($F_{SXR}$ and $F_{HXR}$).  In the thick-target flare model, 
the non-thermal $F_{HXR}$ represents the {\it instantaneous} 
energy deposition rate ($\dot{\mathcal{E}}_e$) by the electron beam precipitating to the chromosphere, 
but the thermal $F_{SXR}$ is proportional to the {\it cumulative} energy deposited,
i.e., the time-integral of $\dot{\mathcal{E}}_e$.
It naturally follows that the temporal derivative of the SXR flux, 
$\dot{F}_{SXR}$, should be related to $F_{HXR}$.

The simplest test of the Neupert effect is usually carried out by plotting 
$\dot{F}_{SXR}$ and $F_{HXR}$ in some energy band. There are many reasons why a simple
linear relationship would not be the case here. The first and most important is
that $\dot{\mathcal{E}}_e$ is related to $F_{HXR}$ through the bremsstrahlung yield
function $Y$ ($F_{HXR} = \dot{\mathcal{E}}_e Y$) which is not a constant and
depends on the spectrum of the electrons or HXRs (see e.g. Petrosian 1973).
Here the most crucial factor is the low energy cutoff ($E_1$) of the non-thermal electrons,
but the spectral index also plays some role. The total yield of all the bremsstrahlung photons
produced by a power law spectrum of electrons with energies above $E_1$ (in units of 511 keV) is
  \begin{equation}
    Y_{total}= \frac{16}{3} \left(\frac{\alpha}{4\pi \ln \Lambda}\right) E_1 
               \left(\frac{\delta-2}{\delta-3}\right),
  \label{Ytotal}\end{equation}
and the yield of the photons whose energies are greater than $E_1$ is 
  \begin{equation}
    Y_{E_1} = \frac{16}{3} \left(\frac{\alpha}{4\pi \ln \Lambda}\right) E_1	
              \left(\frac{2}{\delta-1}\right)^2 \left(\frac{1}{\delta-3}\right),
  \label{Y}\end{equation}
where $\alpha=1/137$, $\ln \Lambda=20$ is the Coulomb logarithm, and
$\delta$ is the spectral index of the power-law electron flux.
As shown in Figure \ref{time}{\it f} both the low energy
cutoff and spectral index of the non-thermal emission vary during the pulse, 
indicating variations in the electron spectrum and thus breaking 
the linearity of the SXR-HXR relationship. Other factors which can also produce
further deviations are energy deposition by protons (and other ions),
by conduction and other possible ways of dissipation of energy
than simply heating and evaporating the chromospheric plasma by non-thermal electrons.
A detailed treatment of the problem requires solutions of the combined transport
and HD equations, which is beyond the scope of this paper. \citet{Vero05}
by inclusion of some of these effects in an approximate way found that the expected relationship
was mostly not present in several {\it RHESSI} flares. Finally one must include
the fact that the chromospheric response of SXR emission will be delayed
by tens of seconds depending on the sound travel time (and its variation) and 
other factors. 

The flare under study has shown no indication of gamma-ray line emission
which means that the contribution of protons most probably is small.
In the currently most favorable model where the electrons are accelerated
stochastically by turbulence (see e.g. Petrosian \& Liu 2004) the turbulence
can suppress heat conduction during the impulsive phase and possibly also during
the decay phase \citep{Jian06}. Because there does not appear to be large changes
in the shape of the loop during the impulsive phase other energy dissipation
processes such as cooling by expansion may also be negligible. Assuming
these to be the case we have performed the Neupert effect test in two ways, 
the first being the common practice of examining the relation between 
$\dot{F}_{SXR}$ and $F_{HXR}$. We then examine the relation between
$\dot{\mathcal{E}}_e$ and $\dot{F}_{SXR}$ by taking into account the 
variation of the bremsstrahlung yield.

1. The temporal derivatives of the fluxes of the two {\it GOES} channels are shown in 
the {\it bottom panel} of Figure \ref{lightcv}. As evident, during the 
rising portion of the {\it GOES} fluxes both channels' derivatives
indeed match the first pulse of the {\it RHESSI} HXR light curves ($> 25$ keV), 
but not during the second weaker pulse (where the 1-8 {\AA} derivative 
shows some instrumental artifacts).
This may be due to the fact that the Neupert effect of the second pulse is overwhelmed
by the cooling of the hot plasma produced during the first stronger pulse.
Nevertheless, the SXR light curves 
(of both {\it GOES} and {\it RHESSI}) exhibit slightly slower decay rate than expected
from the first pulse alone.  This most likely is the signature of the energy input by the second
pulse, which slows down the decay of the first pulse alone.  

We note in passing that the SXR light curves start rising several minutes prior to the onset of 
the HXR impulsive phase.  This is an indication of preheating of the plasma before production
of a significant number of suprathermal electrons.
The 6-12 keV curve rises faster than the {\it GOES} curves at 
lower photon energies, which is consistent with the picture that the primary energy release by 
reconnection occurs high in the corona where the relatively hotter plasma is heated 
before significant acceleration of electrons (as suggested in Petrosian \& Liu 2004), 
and before transport of energy (by accelerated electrons or conduction)
down the flare loop to lower atmosphere where cooler plasmas are heated subsequently
and produce the {\it GOES} flux. On the other hand, 
the increase of the SXR flux at the beginning is dominated by the southern loop,
which shows little evidence of chromospheric evaporation. The phenomenon therefore may 
be a unique feature of this flare.

To quantify the SXR-HXR relationship, we cross-correlated the {\it RHESSI} 30-50 keV photon energy flux
($F_{30-50}$, Figure \ref{Neupert}{\it a}) and the derivative of the {\it GOES} low 
energy channel flux ($\dot{F}_{SXR}$, Figure \ref{Neupert}{\it c}) in the 
SXR rising phase (04:58:00-04:59:51 UT).
The resulting Spearman rank-order correlation coefficient (see Figure \ref{Neupert}{\it f}), an 
indicator of an either linear or nonlinear correlation, shows a single hump with a maximum value 
of 0.91 (corresponding to a significance of $\sim$$10^{-13}$) at a time lag of 12 s. 
This suggests a delay of $\dot{F}_{SXR}$ relative to $F_{30-50}$, which is expected 
given the finite hydrodynamic response time (on the order of the 	
sound travel time of $\sim$20 s for a loop size of $\sim$$10^9$ cm 
and $T \sim 10^7$ K) required for redistribution of the deposited energy.
Such a delay is evident in the numerical simulations of \citet{Li93} who in addition found 
that the density enhancement contributes more to the total SXR emissivity 
than the temperature increase for longer duration ($\geq 30$ s) HXR bursts during the decay phase. 
In Figure \ref{Neupert}{\it d}, we plot the two quantities with the {\it GOES} 
derivative shifted backward by 12 s to compensate the lag of their correlation.  
A linear regression (dotted line) gives  
   $ F_{30-50}= (1.95\pm 0.15) \dot{F}_{SXR} - (3.68\pm 0.48)$ 
with an adjusted coefficient of determination (or so-called R squared) $R_{adj}^2=0.81$ 
close to 1 suggesting a good linear correlation.


2. We also carried out the same analysis for the electron energy power $\dot{\mathcal{E}}_e$, 
assuming a thick-target model of power-law electrons with a low-energy cutoff of $E_1=25$ keV. 
We first obtained the energy flux of all the photons with energies 
greater than $E_1$, $F_{E_1}$, from the 30-50 keV photon energy flux $F_{30-50}$: 
  \begin{equation}
    F_{E_1} = \int _{E_1}^{\infty} J(E) E dE  = F_{30-50} 
              \frac {E_1^{-\gamma+2}} { 30^{-\gamma+2} - 50^{-\gamma+2}}, 
  \label{factor}\end{equation}
where $J(E) \propto E^{-\gamma}$ is the photon flux distribution at the Sun 
(photons keV$^{-1}$ s$^{-1}$) which is obtained
from spectrum fitting (see \S \ref{spec}) and assumed to extend to infinity in energy space. 
We then calculated the power of the electrons by  	
  \begin{equation}
    \dot{\mathcal{E}}_e=F_{E_1}/Y_{E_1},
  \label{edot}\end{equation}
where the bremsstrahlung yield $Y_{E_1}$ is given by equation (\ref{Y})%
  \footnote{We used more accurate results from numerical integration of equation (29)
  in \citet{Petr73} rather than the approximate equation (\ref{Y}) here.
  However, one can still use equation (\ref{Y}) with a simple correction factor of
  $0.0728 \times (\delta-4) + 1$ in the range $4 \leq \delta \leq 9$ 
  to achieve an accuracy of $\lesssim 1\%$.}.  
The resulting $\dot{\mathcal{E}}_e$ is plotted versus time 
and {\it GOES} derivative in {\it panels b} and {\it e} of Figure \ref{Neupert} respectively.
The dotted line in {\it panel e} shows a linear fit ($R_{adj}^2=0.49$) to the data:
   $ \dot{\mathcal{E}}_e = (0.65\pm 0.11) \dot{F}_{SXR} + (1.88\pm 0.34). $ 
The corresponding Spearman correlation coefficient has a peak value of
0.78 (significance $\sim$$10^{-8}$) at a time lag of 3 s (Figure \ref{Neupert}{\it f}). 
As evident, $\dot{\mathcal{E}}_e$ yields no better correlation with $\dot{F}_{SXR}$ 
than $F_{30-50}$, which is similar to the conclusion reached by \citet{Vero05}.
During the HXR decay phase (after 04:59:20 UT) the spectrum becomes softer 
($\gamma>5$) and $\dot{\mathcal E}_e$ decreases much slower than $F_{30-50}$ 
since the bremsstrahlung yield (equation [\ref{Y}]) decreases with the spectral index.
As noted above for these high spectral indexes, 
the emission might be thermal rather than non-thermal. The inferred electron power 
is thus highly uncertain for these times.

As stated earlier, the total energy of the non-thermal electrons is very sensitive
to the low-energy cutoff $E_1$ which is generally not well determined
(cf. Sui et al. 2005).	
We thus set $E_1$ as a free parameter and repeated the above calculation for 
different values of $E_1$ (ranging from 15 to 28 keV). 
We find that, as expected, the temporal $\dot{\mathcal{E}}_e$-$\dot{F}_{SXR}$ relationship	
highly depends on $E_1$. For a small $E_1$ ($\lesssim 20$ keV), $\dot{\mathcal{E}}_e$ keeps rising till
$\sim$04:59:50 UT (near the bottom of the $F_{30-50}$ light curve), which makes the 
$\dot{\mathcal{E}}_e$-$\dot{F}_{SXR}$ correlation completely disappear. On the other hand,
for a large $E_1$ ($> 20$ keV) the correlation is generally good during the
impulsive pulse (through 04:59:10 UT) and the larger $E_1$ the better the correlation. 
This is because the conversion factor 
$ E_1^{-\gamma+2} / ( 30^{-\gamma+2} - 50^{-\gamma+2} ) $
in equation (\ref{factor}) is an increasing (decreasing) function of
the photon spectral index $\gamma$ if $E_1$ is sufficiently small (large). 
For a small $E_1$, for example, the photon energy flux $F_{E_1}$ may have a somewhat large value 
in the valley of the $F_{30-50}$ light curve when $\gamma$ is high. 
In addition, during this time interval the bremsstrahlung yield $Y_{E_1}$ becomes small since 
$\delta$ is large (see equation [\ref{Y}]) and consequently
this may result in a very large $\dot{\mathcal{E}}_e$ by equation (\ref{edot}).

As to the magnitude of the energy flux of non-thermal electrons, 
\citet{Fish85a} in their HD simulations found that the dynamics of the flare loop plasma 
is very sensitive to its value. For a low energy\
flux ($\leq 10^{10}$ ergs cm$^{-2}$ s$^{-1}$), the up-flow velocity of the evaporating plasma 
is $\sim$10's of km s$^{-1}$; for a high energy flux ($\geq 3 \times 10^{10}$ ergs cm$^{-2}$ s$^{-1}$),
a maximum up-flow velocity of $\sim$100's of km s$^{-1}$ can be produced. 
For the flare under study, we estimate the area of the cross-section of the loop 
to be $ A_{loop} \lesssim 1.6 \times 10^{18}$ cm$^{2}$,
where the upper limit corresponds to the loop width determined by the 5\% level
in Figure \ref{loop-model}{\it b}. We read the maximum electron power of 
   $\dot{\mathcal{E}}_{e, max} = 9.8 \times 10^{28}$ ergs s$^{-1}$
from Figure \ref{Neupert}{\it b}, which is then divided by $2 A_{loop}$ (assuming a filling factor
of unity) to yield the corresponding electron energy flux: 
  $f_{e, max} \gtrsim 3.1 \times 10^{10}$ ergs cm$^{-2}$ s$^{-1}$.
The source velocity on the order of a few hundreds of km s$^{-1}$ estimated
in \S \ref{morp} is consistent with that predicted by \citet{Fish85a}.
For comparison, we note that \citet{Mill06} from {\it RHESSI} data 
also obtained an energy flux of $\geq 4 \times 10^{10}$ ergs cm$^{-2}$ s$^{-1}$ for an M2.2 flare during
which an up-flow velocity of $\sim$230 km s$^{-1}$ was inferred from		
simultaneous co-spatial {\it SoHO}/CDS Doppler observations.		
 
In summary, the {\it GOES} SXR flux derivative $\dot{F}_{SXR}$ exhibits a Neupert-type
linear correlation with the {\it RHESSI} HXR flux $F_{30-50}$ during the first HXR pulse. 
However, unexpectedly, the correlation between the electron power 
$\dot{\mathcal{E}}_e$ and $\dot{F}_{SXR}$
is not well established based on the simple analysis presented here, 	
which suggests that a full HD treatment is needed
to investigate the chromospheric evaporation phenomenon (see discussions in \S \ref{conclusion}).


\section{Loop Density Derivation}
\label{dens}

\begin{figure}[thb]  
    \epsscale{0.32}  
    \plotone{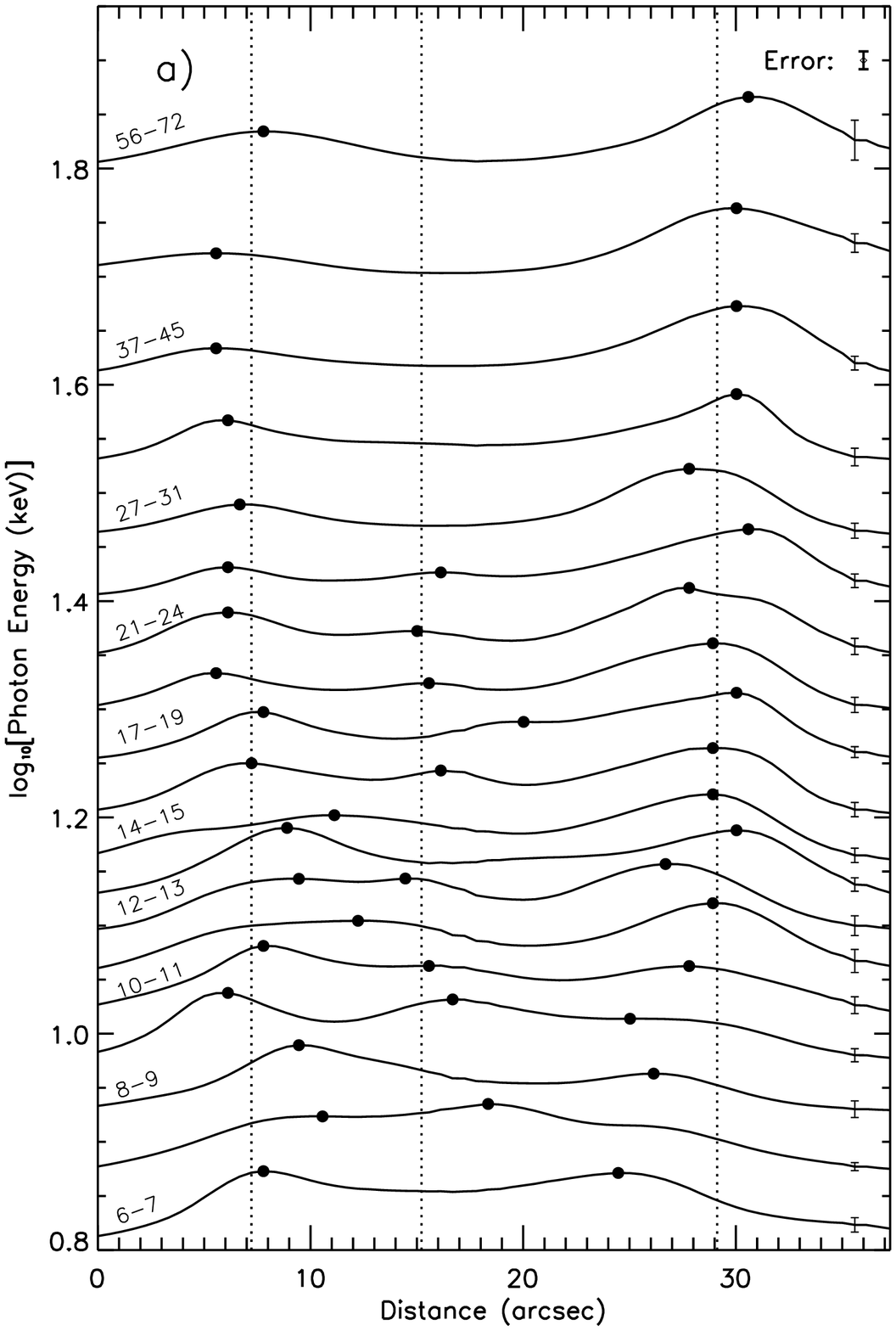}
    \plotone{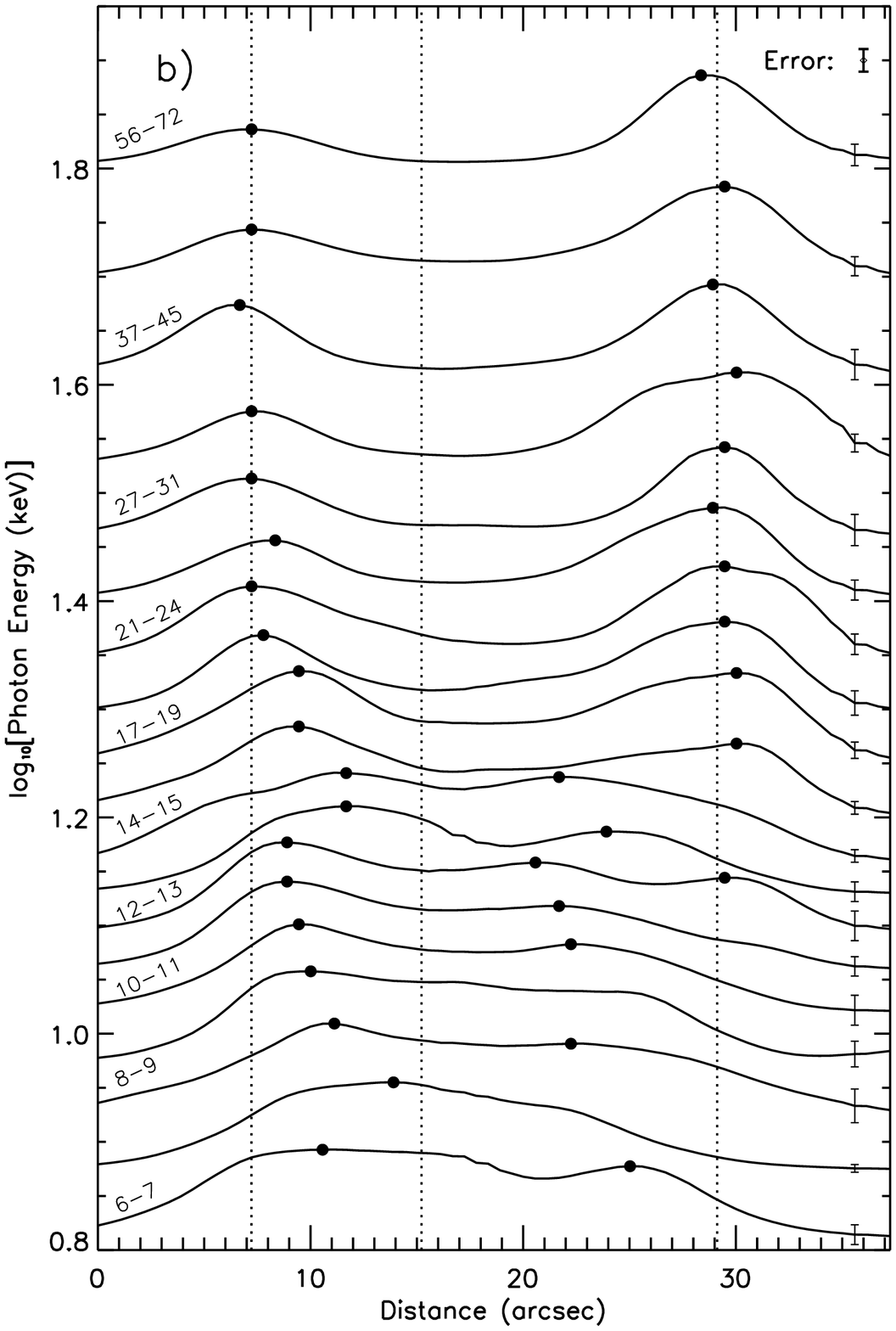}
    \plotone{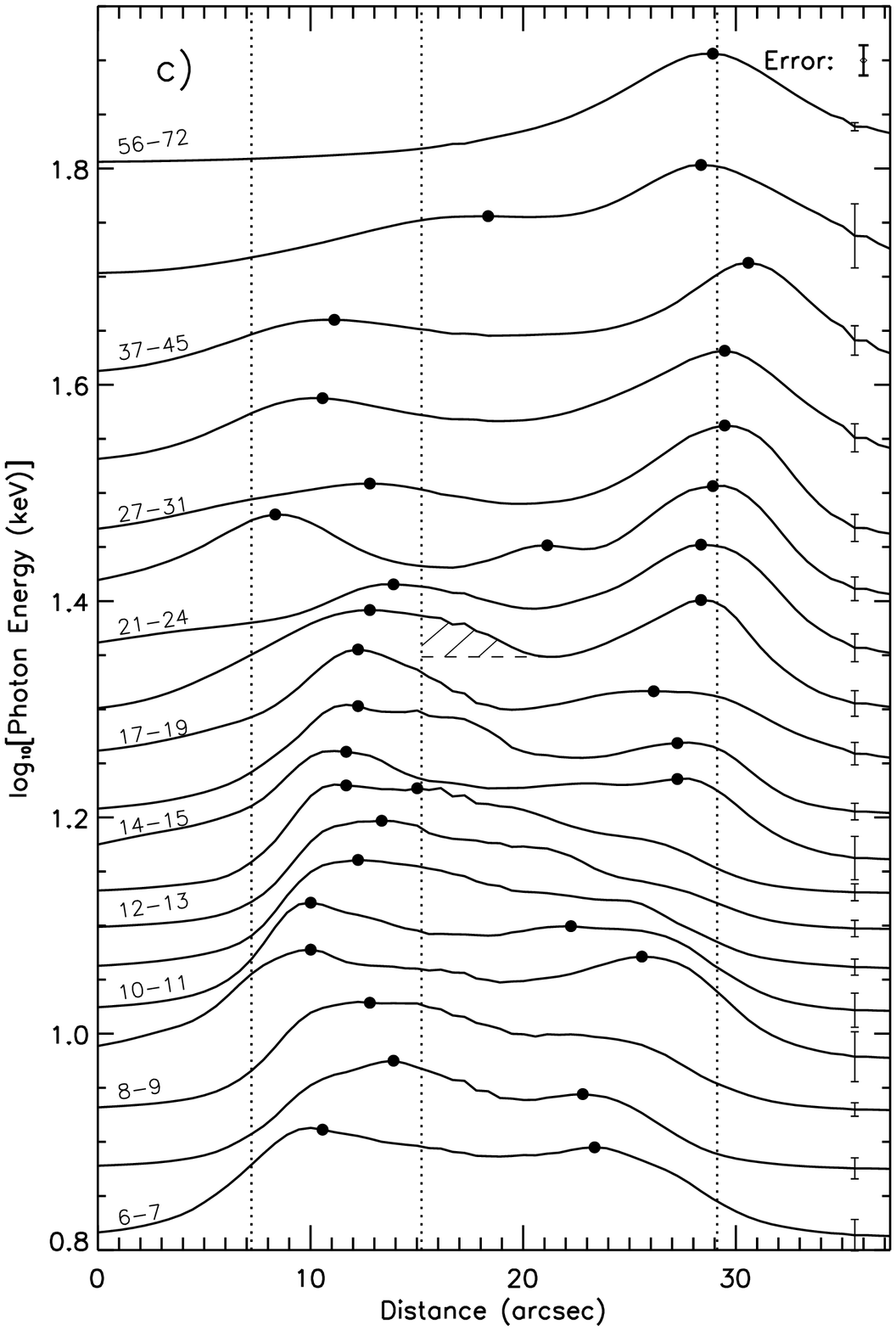}
    \caption{
    ({\it a}) {\it left}: Brightness profiles (obtained in the same way as Figure \ref{prof_3e}) 	
    at different energies for the time interval of 04:58:00-24 UT. 
    The vertical axis is the average photon energy (in logarithmic scale) of the 
    energy band for the profile. Representative energy bands (in keV) are labeled above the
    corresponding profiles. The vertical dotted lines are the same as in 
    Figures \ref{loop-model} and \ref{prof_3e}.		
      ({\it b}) {\it middle} and ({\it c}) {\it right}: Same as {\it panel a} but for 04:58:24-48 and
    04:58:48-59:12 UT, respectively. The error bars are the uncertainties of the corresponding profiles.
    The overall uncertainties, as indicated by the stand-alone error bar in the upper-right 
    corner of each panel (note different scales, similar to Figure \ref{prof_3e}), 
    are 14\%, 13\%, and 14\%, respectively.
    The hatched region in {\it panel c} represents the LT emission (19-21~keV) being removed for
    deriving the density distribution in Figure \ref{dens-fig} (see text).
    }
    \label{prof_24s}	
  \end{figure}

\begin{figure}[thb]  
\epsscale{0.8}
\plotone{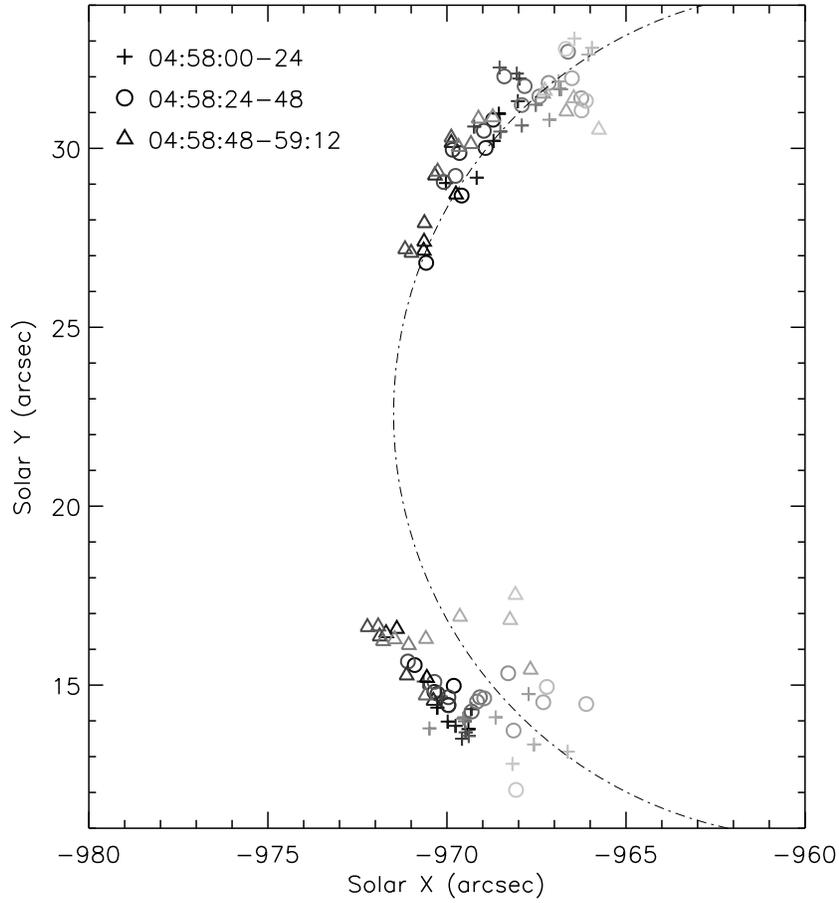}
\caption{Centroids of the northern and southern halves of the loop at different energies
for the three 24-second time intervals (same as those in Figures 
\ref{prof_24s}{\it a}-\ref{prof_24s}{\it c}).	
Energy increases from dark to light-grey symbols. The dot-dashed line marks the central arc of 
the model loop (same as in Figure \ref{loop-model}{\it a}).}
\label{centroids}
\end{figure}

\begin{figure}[thb]  
\epsscale{0.8}
\plotone{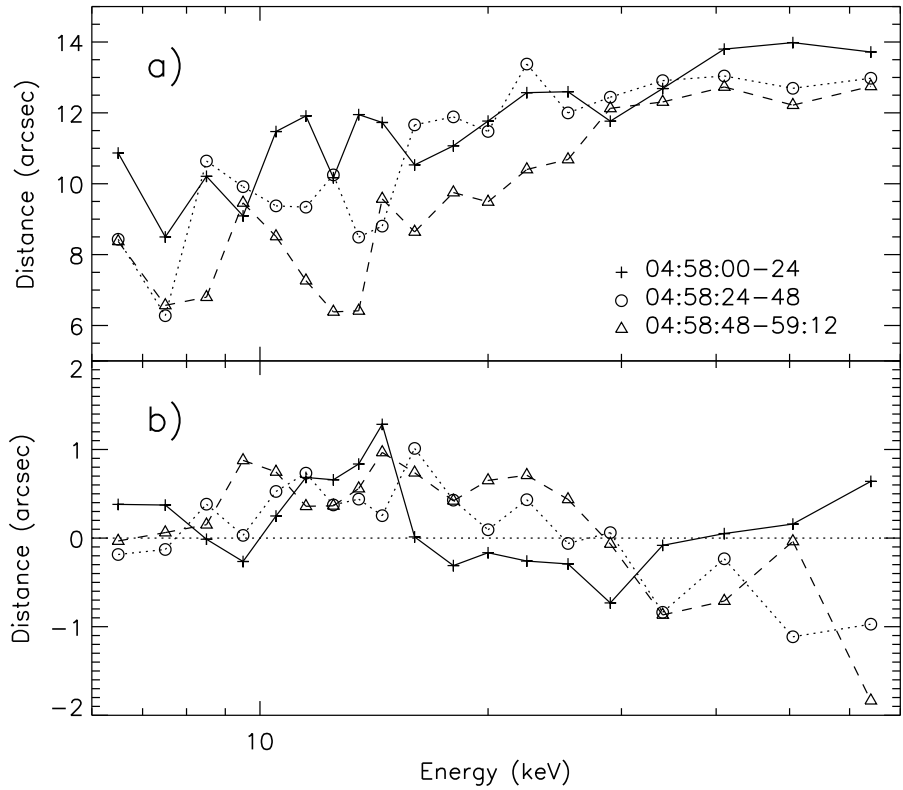}
\caption{Positions of the northern centroids projected along ({\it panel a}) and perpendicular 
to ({\it panel b}, note the different scales) the central arc (the line in Figure \ref{centroids}) 
of the loop. The distance in {\it panel a} is calculated from the average LT position, 
as shown in Figure \ref{loop-model}{\it a}.}
\label{centr-dist}
\end{figure}

\begin{figure}[thb]  
\epsscale{0.70}		
\plotone{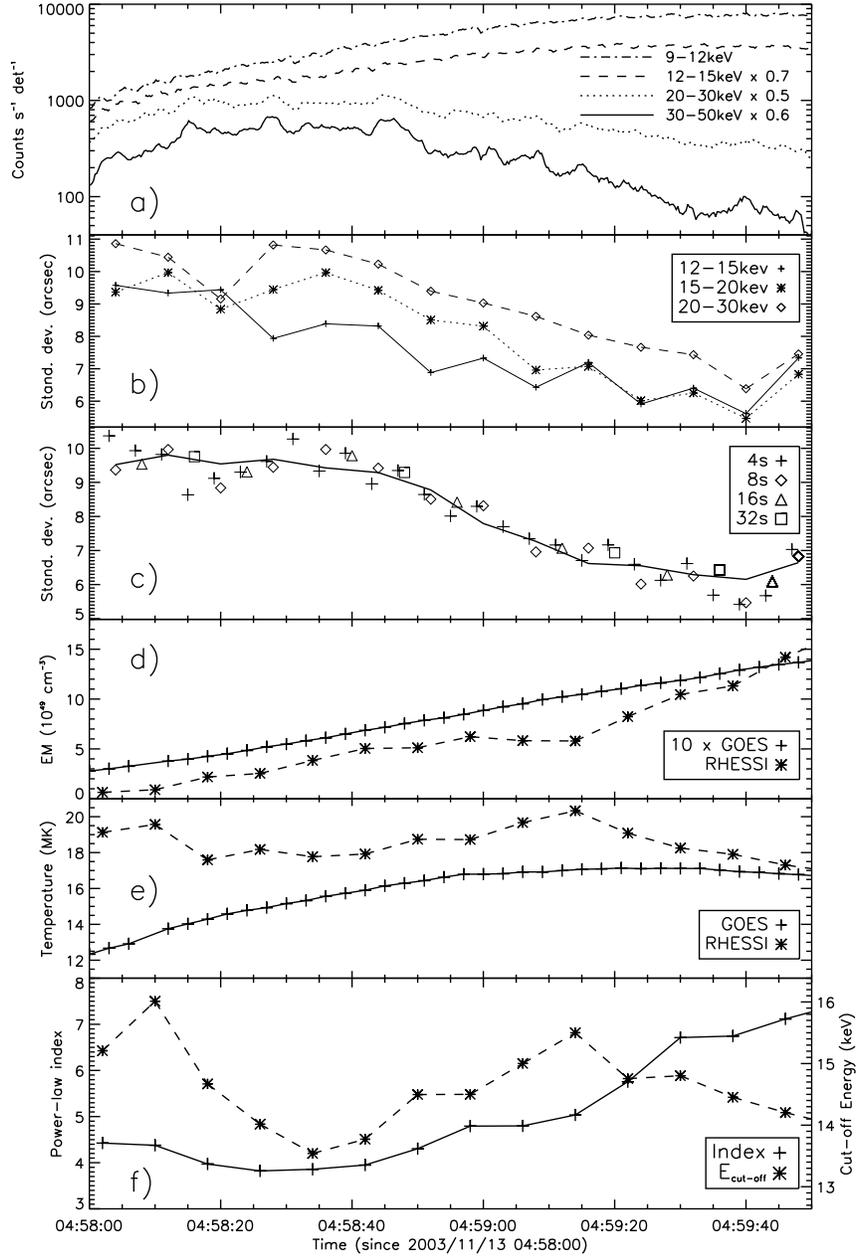}
\caption{({\it a}) {\it RHESSI} light curves (demodulated to remove artificial periodicity
caused by the spacecraft spin).
  ({\it b}) Evolution of the standard deviation of the brightness profiles along the loop
in three different energy bands obtained from CLEAN images.
  ({\it c}) Same as {\it b} but in the 15-20 keV band with different integration time intervals
indicated in the legend. The solid curve denotes the result from PIXON images with an $\sim 8$ 
second integration time interval. 
  ({\it d}) and ({\it e}) Evolution of the emission measure ($10^{49}$ cm$^{-3}$) 
and temperature (MK), respectively,
of the thermal component of the spatially integrated {\it RHESSI} spectrum obtained
from fits to a thermal plus a power law model and from thermal fits to the {\it GOES} spectrum.
The {\it GOES} emission measure is scaled by a factor of 10.
  ({\it f}) The evolution of the power-law index and the low-energy cutoff of 
the {\it RHESSI} power law component.
}
\label{time}
\end{figure}

\begin{figure}[thb]  
\epsscale{1.0}	
\plotone{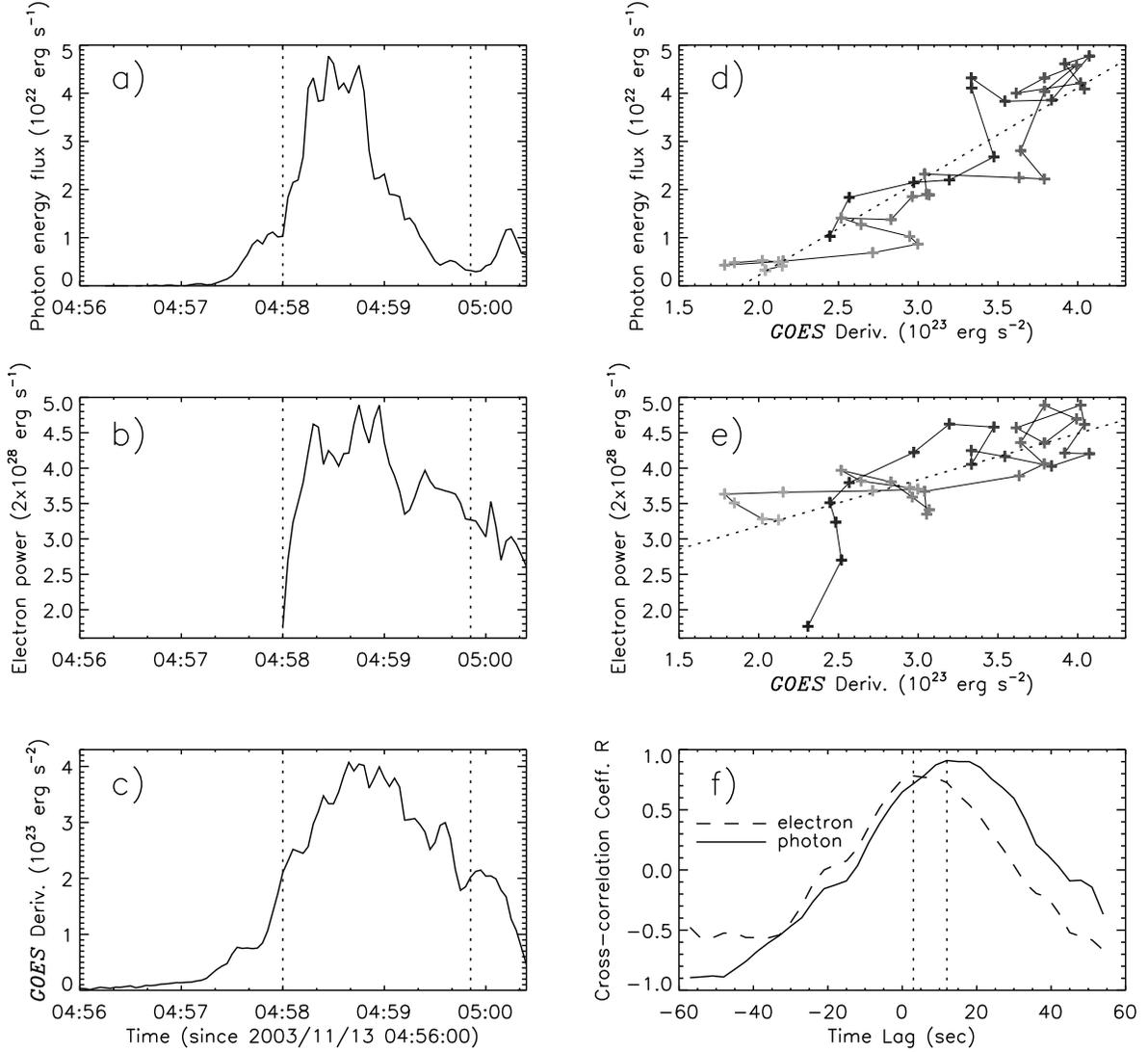}
\caption{({\it a}) The 30-50 keV photon energy flux ($F_{30-50}$) at the Sun inferred from 
the {\it RHESSI} observation at 1 AU assuming isotropic emission.
The two vertical dotted lines outlines the time interval (04:58:00-04:59:51 UT)
used for the cross-correlation analysis (see below).
  ({\it b}) Power ($\dot{\mathcal{E}}_e$) of the power-law electrons 
with a low energy cutoff of 25 keV
inferred from the photon energy flux assuming a thick-target model.
  ({\it c}) Same as {\it panel a} but for the derivative ($\dot{F}_{SXR}$) 
of the {\it GOES} low energy channel (1-8 \AA) flux. 
  ({\it d}) 	
$F_{30-50}$ versus $\dot{F}_{SXR}$
(shifted back in time by 12 s to account for its delay   
 as revealed by the cross-correlation analysis; see {\it panel f})
within the interval of 04:58:00-04:59:51 UT.
The grey scale of the plus signs (connected by the solid lines) from dark to light indicates the 
time sequence.  The dotted line is the best linear fit to the data.		
  ({\it e}) Same as {\it panel d} but for $\dot{\mathcal{E}}_e$	and $\dot{F}_{SXR}$	
which is shifted back by 3 s in time.
  ({\it f}) The Spearman rank-order correlation coefficient $R$ of the photon energy flux (electron power)
and $\dot{F}_{SXR}$		
plotted as a function of time lag of the latter 
relative to the former.  The dotted lines mark the peak values of $R=0.91$ and 0.78 
at a lag of 12 and 3 s, respectively.
}
\label{Neupert}
\end{figure}

For the 1994 June 20 disk flare, \citet{Silv97} interpreted the moving SXR sources as 
thermal emission from the hot ($\sim 30-50$ MK) plasma evaporated from the chromosphere 
based on the good agreement of the emission measure of the blue-shifted component and that of the 
SXR from the FPs. For the limb flare under study here, Doppler shift measurements are not 
available. Meanwhile, a purely thermal scenario would have difficulties in explaining the 
systematic shift of the centroids towards the FPs with increasing energies up to $\sim$70 keV 
as shown in Figure \ref{centr-dist}. A non-thermal scenario appears more appropriate. 
That is, the apparent HXR FP structure and motions can result from a decrease in the stopping 
distance of the non-thermal electrons with decreasing energy and/or increasing ambient plasma 
density caused by the chromospheric evaporation (as noted earlier in \S \ref{morp}).
One can therefore derive the density distribution along the loop from the corresponding
X-ray emission distributions (e.g., Figure \ref{prof_24s}),		
without any pre-assumed density model \citep[c.f.][]{Asch02}.  
This approach is described as follows.

For a power-law X-ray spectrum produced by an injected power-law electron spectrum, 
\citet{Leac84} obtained a simple empirical relation \citep[also see][\S 2]{Petr99} 
for the X-ray intensity $I(\tau, k)$ per unit 
photon energy $k$ (in units of 511 keV) and column depth $\tau$ (in units of 
$1/[4 \pi r^2_0 \ln \Lambda] = 5 \times 10^{22}$ cm$^{-2}$ for $r_0=2.8 \times 10^{-13}$ cm
and $\ln \Lambda = 20$):
  \begin{equation}
    I(\tau, k) = A \left(\frac{\delta}{2}-1\right) \left(\frac{k+1}{k^{2+\gamma}}\right) \left(1 + 
                 \tau \frac{k+1}{k^2}\right)^{-\delta/2},
  \end{equation}
where $\gamma$ and $\delta$ $(= \gamma+0.7)$ are the photon and electron spectral indexes, respectively, 
$A$ is a constant normalization factor, and $d \tau= n ds$, 
where $s$ is the distance measured from the injection site.
This equation quantifies the dependence of the emission profile (or source morphology)
on the electron spectral index and column depth. 
In general, when $\delta$ decreases (spectrum hardening), 
the intensity at a given photon energy rises (drops) at large (small) $\tau$'s and thus the
emission centroid shifts to larger $\tau$'s. This is expected because for a harder spectrum,
there are relatively more high energy electrons that can penetrate to larger column depths
and produce relatively more bremsstrahlung photons there. The opposite will happen when the spectrum 
becomes softer. During the impulsive peak showing a soft-hard-soft behavior 
(see \S \ref{spec}), one would expect that the emission centroids shift first away from and 
then back toward the LT (if the density in the loop stays constant).  
Knowing the spectral index, the emission profile can therefore yield critical 
information about the density variation in both space and time.


To compare the above empirical relation with observations, we first integrate $I(\tau, k)$ 
over an energy range $[k_1, k_2]$,
  \begin{equation}
    J(\tau; k_1, k_2)		
      = \int _{k_1}^{k_2} A \left(\frac{\delta}{2}-1\right) \left(\frac{k+1}{k^{2+\gamma}}\right) 
        \left(1 + \tau \frac{k+1}{k^2}\right)^{-\delta/2} dk,
  \end{equation}
and then integrate $J(\tau; k_1, k_2)$ over $\tau$ to obtain the cumulative emission, 
  \begin{equation}
    F(\tau; k_1, k_2) = \int _{0}^{\tau} J(\tau; k_1, k_2) d\tau 
         = \frac {1-\gamma} { k_2^{1-\gamma} - k_1^{1-\gamma}} 
           \int _{k_1}^{k_2} \left[1 - \left(1 + \tau \frac{k+1}{k^2}\right)^{1- \delta/2}\right] 
           k^{-\gamma} dk,
  \label{Ftau}\end{equation}
where we have chosen
  \begin{equation}
     A = \left(\int_{k_1}^{k_2} k^{-\gamma}dk\right)^{-1} = \frac {1-\gamma} { k_2^{1-\gamma} - 
         k_1^{1-\gamma}} ,
  \end{equation}
so that $F(\tau=\infty; k_1, k_2)=1$.
Comparison of $F(\tau; k_1, k_2)$ with the observed emission profiles gives
the column depth $\tau(s)$ whose derivative with respect to $s$ then gives the density profile along 
the loop.

Specifically for this flare, we assume that the non-thermal electrons are injected at the LT 
indicated by the middle vertical dotted line in Figure \ref{prof_24s}	
and denote the profile to the right-hand side of this line (i.e., along the 
northern half of the loop) as $J_{obs}(s; k_1, k_2)$, where $[k_1, k_2]$ is the energy band of 
the profile.  The observed cumulative emission is then given by,
  \begin{equation}
    F_{obs}(s; k_1, k_2) = \int _0^s J_{obs}(s; k_1, k_2) ds /  \int _0^{s_{max}} J_{obs}(s; k_1, k_2) ds,
  \end{equation}
where $s_{max}$ (corresponding to $\tau=\infty$) is the maximum distance considered 
and $F_{obs}(s; k_1, k_2)$ has been properly normalized. 
Then $\tau = \tau(s; k_1, k_2)$ can be obtained by inverting  
  \begin{equation}
    F(\tau; k_1, k_2) =F_{obs}(s; k_1, k_2),
  \end{equation}
where the integration over $k$ in equation (\ref{Ftau}) can be calculated numerically.

\begin{figure}[thb]  
\epsscale{0.9}
\plotone{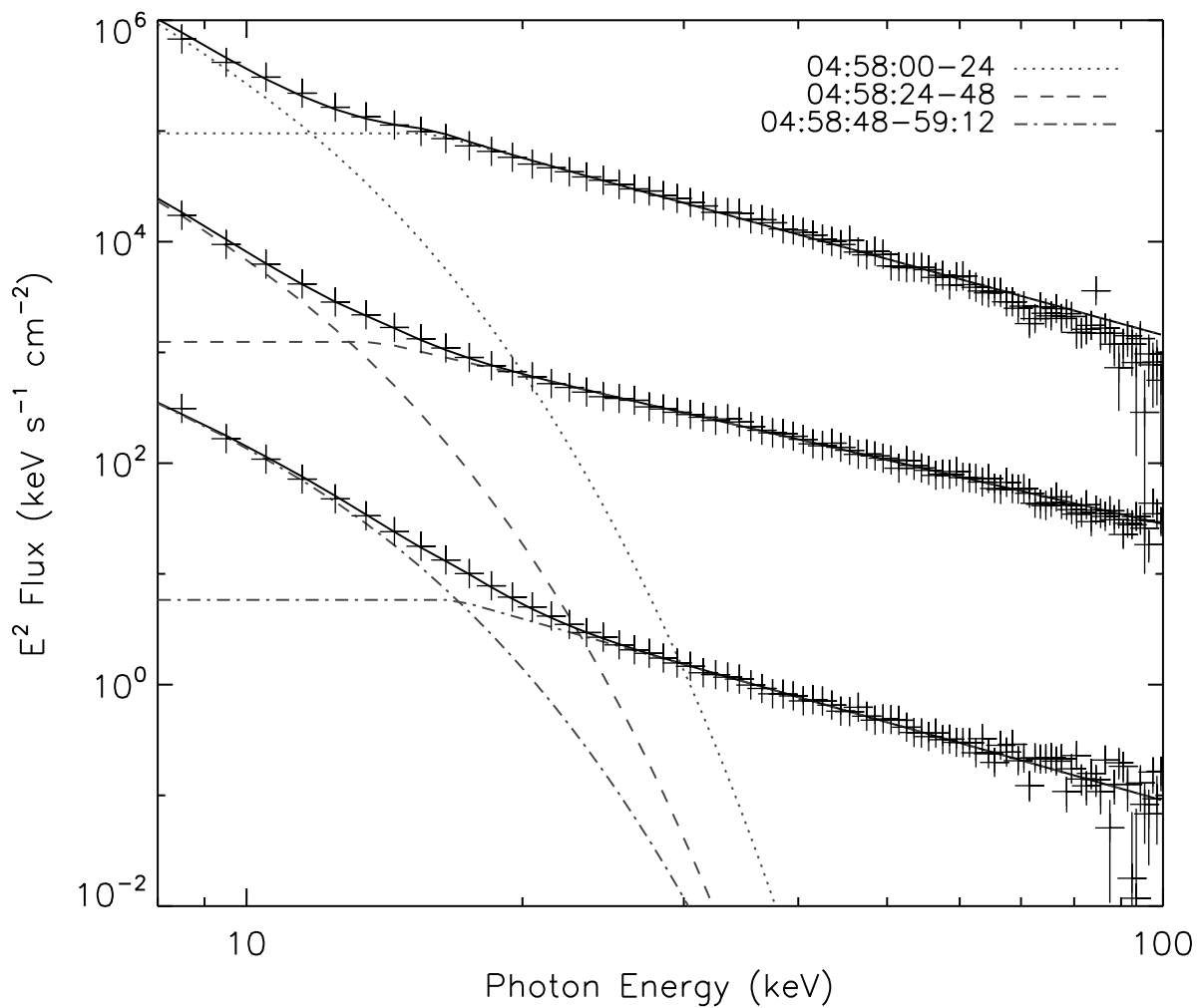}
\caption{The $\nu F_{\nu}$ spatially integrated spectra for the three 24 s time intervals
as indicated in the legend. From the top to the bottom, 
the 2nd and 3rd spectra are shifted downward by 2 and 4 decades, respectively. The broken
lines are the thermal and power-law components of the fits to the data, and the solid lines 
are the sum of the two components. The thermal and power-law components intersect at
about 12, 13, and 17 keV, respectively for the three intervals, above which the power-law 
component dominates.
}
\label{spec_24s}
\end{figure}

It should noted that, however, not all the profiles in Figure \ref{prof_24s} are suitable
for this calculation, because low energy emission is dominated by a thermal component
especially in the LT region and at later times. We thus restrict ourselves to the energy
ranges of 12-72, 13-72, and 17-72 keV, respectively for the three 24 s intervals. 
The lower bound is the energy above which the power-law component dominates over the 
thermal component, determined from fits to the spatially integrated spectrum for each 
interval as shown in Figure \ref{spec_24s}. 
Within these energy ranges, separate leg or FP sources rather than a single 
LT source can be identified in the corresponding image, which is morphologically 
consistent with the non-thermal nature of emission assumed here. 
To further minimize the contamination of the thermal emission in our analysis, 
we have excluded the LT portion of the emission profile in excess of the 
lowest local minimum (if it exists) between the LT and leg (or FP) sources. 
An example of this exclusion is illustrated with the hatched 
region in Figure \ref{prof_24s}{\it c} for the 19-21~keV profile. 
This was done by simply replacing the profile values between the LT 
and the local minimum positions with the value at the minimum.

\begin{figure}[thb]  
\epsscale{0.8}
\plotone{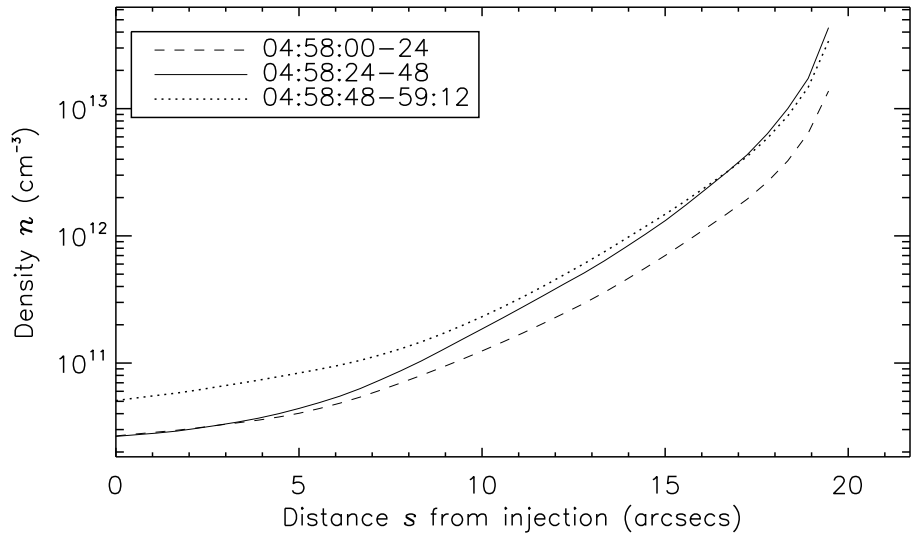}
\caption{Averaged density profiles along the loop inferred from the HXR brightness 
profiles during the three time intervals.}
\label{dens-fig}
\end{figure}

We calculated $\tau(s; k_1, k_2)$ for every emission profile within the energy 
ranges mentioned above for the three intervals in Figure \ref{prof_24s}, 
with the photon indexes, $\gamma=$ 4.46, 3.97, and 4.23 respectively.
From the geometric mean of the column depths obtained at different energies, $\bar\tau$, we derived 
the density profile $n(s) = {\rm d}\bar\tau(s)/{\rm d}s$ for each time interval. The results are shown 
in Figure \ref{dens-fig}, where we shall bear in mind that attention should be paid to
the overall trend rather than the details of the density profile and its variation, 
because the profile here only spans about 3 times the resolution ($\sim 7\arcsec$)
and thus is smoothed, making neighboring points not independent. 
As can be seen, between the 1st and 2nd intervals, the density increases dramatically
in the lower part of the loop, while the density near the LT remains essentially 
unchanged. The density enhancement then shifts to the LT from the 2nd to the 3rd interval. 
This indicates a mass flow from the chromosphere to the LT. 
The density in the whole loop is about doubled over the three intervals, which is
roughly consistent with the density change inferred from the emission measure%
  \footnote{From 04:58:12 through 04:59:00 UT, the {\it RHESSI} ({\it GOES}) emission measure
  rises by a factor of 5.3 (2.3) which translate to an increase of the density 
  by a factor of 2.3 (1.5), assuming a constant volume.} 
(see Figure \ref{time}{\it d}). 
These results are again compatible with the chromospheric 
evaporation picture discussed in \S \ref{morp}.


\section{Conclusion and Discussion} 
\label{conclusion}

We have presented in this paper a study of {\it RHESSI} images and spectra of the 2003 November 13
M1.7 flare.  {\it RHESSI}'s superior capabilities reveal great details of the HXR source 
morphology at different energies and its evolution during the impulsive phase. 
The main findings of this paper are as follows.
 (1) The energy dependent source morphology in general shows a gradual shift of emission from the 
LT to the FPs with increasing energies. Over some short integration intervals emission from the loop
legs may dominate at intermediate energies. 
 (2)	
The emission centroids move toward the LT along the loop during the rising and plateau portions 
of the impulsive phase.
This motion starts at low energies and proceeds to high energies.
We estimate the mean velocity of the motion to be hundreds of km s$^{-1}$, which agrees 
with the prediction of the hydrodynamic simulations by \citet{Fish85a}. There are also shorter time scale
variations that imply much higher velocities ($\sim$$10^3$ km s$^{-1}$) but we are not certain 
if they are real because of instrumental limitations.
 (3) Fits to the spatially integrated {\it RHESSI} spectra with a thermal plus a power law model 
reveal a continuous increase of the emission measure while the temperature does not change significantly.
The {\it GOES} data show a similar trend of the $EM$ 	
but a gradual increase of the temperature.
 (4) The time derivative of the {\it GOES} SXR flux is correlated with the
{\it RHESSI} HXR flux with a peak correlation coefficient of 0.91 at a delay of 12 s in 
agreement with the general trend expected from the Neupert effect. 
However, the correlation between the electron power and the {\it GOES} derivative
is no better than the SXR-HXR correlation.
 (5) From the observed brightness profiles we derive the 
spatial and temporal variation of the plasma density in the loop,	
assuming a non-thermal thick-target bremsstrahlung model. 		
We find a continuous increase of the density, starting first at the FPs and legs and then reaching
to the LT.
All these results fit into a picture of continuous chromospheric evaporation 
caused by the deposition of energy of electrons accelerated during the
impulsive phase.


Several of the new features of this event 
(such as the leg emission at intermediate energies) may be common 
to many solar flares. Expanding the sample of flares of this kind will be very helpful 
to understand the underlying physical processes. 
The new findings reside near the 
limit of {\it RHESSI}'s current temporal, spatial, and spectral resolution.
As advanced imaging spectroscopy capabilities are being developed and spatial resolution
is being improved in the {\it RHESSI} software \citep{Hurf02}, 
it will be critical to obtain the spatially resolved photon spectrum along
the loop. This will yield incisive clues to the nature of the moving X-ray sources
and relevant energy transport mechanisms
and will be useful to check the reality of the short time scale variations.

There are several important questions that 
need to be further addressed in future observational and theoretical investigations: 
  (1) What is the nature of the moving X-ray sources? Could they be characterized as thermal 
emission from the evaporated hot plasma or non-thermal emission from the precipitating electrons, 
or a mixture of both? Could they be related to MHD waves or evaporation fronts?
  (2) What are the roles of different heating agents of the chromosphere, 
i.e., electron beams, thermal conduction, and/or direct heating by turbulence 
or plasma waves during the impulsive phase? 

We have pointed out to some of the many physical processes that come into play 
in answering such questions. Here we describe possible directions for future theoretical studies. 
We have shown that a more physical based test of the Neupert effect between
the electron power and SXR flux derivative does not reveal a
better correlation than the usual HXR vs SXR derivative correlation. 
Although the observed source velocity agrees with those of HD simulations, 
there are some features that current simulations have not addressed.
To answer these questions requires an updated numerical calculation where one combines the model 
of particle acceleration and transport with the HD simulation of 
the atmospheric response to energy deposition to form a unified picture
of solar flares. For example, one can use the output electron spectrum from the 
stochastic particle acceleration model \citep{Hami92, Mill97, Park97, Petr04}
as the input to the transport and HD codes, 
rather than simply assume a power-law electron spectrum as previous HD simulations. 
Such a study can shed light on the relative importance of particle beams 
and thermal conduction in evaporating chromospheric plasma and the roles that MHD waves may play 
in heating the flaring plasma, in particular, 
addressing our tentative observation of the fast source motion
which suggests possible presence of MHD waves in the flare loop. 
A better understanding of their propagation, damping, and excitation mechanisms is necessary for 
uncovering the energy release process during flares. This is particularly true in the context of the 
stochastic particle acceleration model.



\acknowledgments{
This work was supported by NASA grants NAG5-12111, NAG5 11918-1, and NSF grant ATM-0312344.
The work performed by S.L.\ was funded in part under the auspices of the U.S.\ Dept.\ of Energy, 
and supported by its contract W-7405-ENG-36 to Los Alamos National Laboratory.
W.L.\ would like to acknowledge generous computing support from the 
Stanford Solar Physics Group led by P. Scherrer.
We thank the referee for valuable suggestions that helped improve this paper.
We are indebted to G. Hurford for helpful discussions on imaging techniques
and B. Dennis for suggestions on studying the Neupert effect. We are also grateful to
R. Schwartz, T. Metcalf, D. Smith, J. McTiernan, K. Tolbert, S. Krucker and
other members of the {\it RHESSI} team for help of various kinds. 

}


{}



\clearpage

\begin{thebibliography}{}


\bibitem[Allred et al.(2005)]{Allr05}
{Allred, J. C., Hawley, S. L., Abbett, W. P., \& Carlsson, M. 2005, 630, 573}

\bibitem[Antonucci et al.(1990)]{Anto90}
Antonucci, E., Dodero, M. A., \& Martin, R. 1990, \apjs, 73, 147

\bibitem[Antonucci et al.(1982)]{Anto82}{Antonucci, E., et al. 1982, \solphys, 78, 107}

\bibitem[Antonucci et al.(1984)]{Anto84}
{Antonucci, E., Gabriel, A. H., \& Dennis, B. R. 1984, \apj, 287, 917}

\bibitem[Antonucci et al.(1999)]{Anto99}
{Antonucci, E., Alexander, D., Culhane, J. L., de Jager, C., MacNeice, P., Somov, B. V., 
Zarro, D. M. 1999, in The Many Faces of the Sun, 
ed. K. T. Strong et al. (New York: Springer-Verlag), 345}


\bibitem[Aschwanden et al.(2002)]{Asch02}
{Aschwanden, M. J., Brown, J. C., \& Kontar E. P. 2002, \solphys, 210, 383}

\bibitem[Brosius(2003)]{Bros03}{Brosius, J. W. 2003, \apj, 586, 1417}

\bibitem[Brosius \& Philips(2004)]{Bros04}
{Brosius, J. W., \& Philips, K. J. 2004, \apj, 613, 580}


 \bibitem[Chernov et al.(2005)]{Cher05}
 Chernov, G. P., Yan, Y. H., Fu, Q. J., \& Tan, Ch. M. 2005, \aap, 437, 1047

\bibitem[Dennis \& Zarro(1993)]{Denn93}
{Dennis, B. R., \& Zarro, D. M. 1993, \solphys, 146, 177}
 
\bibitem[Dennis et al.(2003)]{Denn03}
{Dennis, B. R., Veronig, A., Schwartz, R. A., Sui, L., Tolbert, A. K., Zarro, D. M.,
 \& {\it RHESSI} Team. 2003, Adv. Space Res., 32, 2459} 

\bibitem[Dere et al.(2000)]{Dere00}{Dere, K. P. et al. 2000, \solphys, 195, 13}

\bibitem[Doschek et al.(1980)]{Dosc80}
{Doschek, G. A., Feldman, U., Kreplin, R. W., \& Cohen, L. 1980, \apj, 239, 725}

\bibitem[Doschek et al.(1994)]{Dosc94}
Doschek, G. A., et al. 1994, \apj, 431, 888

\bibitem[Feldman et al.(1980)]{Feld80}
{Feldman, U., Doschek, G. A., Kreplin, R. W., \& Mariska, J. T. 1980, \apj, 241, 1175}

\bibitem[Fisher et al.(1985a)]{Fish85a}
{Fisher, G. H., Canfield, R. C., \& McClymont, A. N. 1985a, \apj, 289, 414}		

\bibitem[Fisher et al.(1985b)]{Fish85b}
{------. 1985b, \apj, 289, 434}

\bibitem[Gan et al.(1995)]{Gan95}{Gan, W. Q., Cheng, C. C., \& Fang, C. 1995, \apj, 452, 445}

\bibitem[Hamilton \& Petrosian(1992)]{Hami92}
Hamilton, R. J., \& Petrosian, V. 1992, \apj, 398, 350

\bibitem[Holman et al.(2003)]{Holm03}
{Holman, G. D., Sui, L., Schwartz, R. A., \& Emslie, A. G. 2003, \apj, 595, L97}

\bibitem[Hudson(1991)]{Huds91} {Hudson, H. S. 1991, Bull. Am. Astron. Soc., 23, 1064}

\bibitem[Hurford et al.(2002)]{Hurf02}
{Hurford, G. J. et al. 2002, \solphys, 210, 61}

\bibitem[Jiang et al.(2006)]{Jian06}
{Jiang, Y. W., Liu, S., Liu, W., \& Petrosian, V. 2006, \apj, 638, 2}

\bibitem[Leach(1984)]{Leac84}{Leach, J. 1984, Ph.D. Thesis, Stanford University}


\bibitem[Li et al.(1993)]{Li93} {Li, P., Emslie, A. G., Mariska, J. T. 1993, \apj, 417, 313}

\bibitem[Lin et al.(2002)]{Lin02}{Lin, R. P., et al. 2002, \solphys, 210, 3}

\bibitem[Liu et al.(2004)]{Liu04}
{Liu, W., Jiang, Y. W., Liu, S., \& Petrosian, V. 2004, \apj, 611, L53}

\bibitem[Mariska et al.(1989)]{Mari89}
{Mariska, J. T., Emslie, A. G., \& Li, P.  1989, \apj, 341, 1067}

\bibitem[Metcalf et al.(1996)]{Metc96}
{Metcalf, T. R., Hudson, H. S., Kosugi, T., Puetter, R. C., \& Pina, R. K. 1996, \apj, 466, 585}

\bibitem[Metcalf et al.(2005)]{Metc05}
{Metcalf, T. R., Leka, K. D., Mickey, D. L. 2005, \apj, 623, L53}

\bibitem[Miller et al.(1997)]{Mill97}
{Miller, J. A., et al. 1997, \jgr, 102, 14631}

\bibitem[Milligan et al.(2006)]{Mill06}
Milligan, R. O., Gallagher, P. T., Mathioudakis, M., Bloomfield, D. S., 
Keenan, F. P., \& Schwartz, R. A. 2006, \apj, 638, L117

\bibitem[Neupert(1968)]{Neup68}{Neupert, W. M. 1968, \apj, 153, L59}

\bibitem[Park et al.(1997)]{Park97}
Park, B. T., Petrosian, V., \& Schwartz, R. A. 1997, \apj, 489, 358

\bibitem[Petrosian(1973)]{Petr73}{Petrosian, V. 1973, \apj, 186, 291}

\bibitem[Petrosian \& Donaghy(1999)]{Petr99}
{Petrosian, V., \& Donaghy, T. Q. 1999 \apj, 527, 945}

\bibitem[Petrosian \& Liu(2004)]{Petr04}{Petrosian, V. \& Liu, S. 2004, \apj, 610, 550}

\bibitem[Peres \& Reale(1993)]{Pere93}{Peres, G., \& Reale, F. 1993, \aap, 275, L13}

\bibitem[Schmahl \& Hurford(2002)]{Schm02}
{Schmahl, E. J., \& Hurford, G. J. 2002, \solphys, 210, 273}

\bibitem[Silva et al.(1997)]{Silv97}
{Silva, A. V. R., Wang, H., Gary, D. E., Nitta, N., \& Zirin, H. 1997, \apj, 481, 978}

\bibitem[Smith et al.(2002)]{Smit02}
{Smith, D., et al. 2002, \solphys, 210, 33}


\bibitem[Sui et al.(2004)]{Sui04}{Sui, L., Holman, G. D., \& Dennis, B. R. 2004, \apj, 612, 546}

\bibitem[Sui et al.(2005)]{Sui05}{------. 2005, \apj, 626, 1102}

\bibitem[Veronig et al.(2005)]{Vero05} {Veronig, A. M., Brown, J. C., Dennis, B. R.,
 Schwartz, R. A., Sui, L., \& Tolbert, A. K. 2005, \apj, 621, 482}

\bibitem[Veronig et al.(2006)]{Vero06} {Veronig, A. M., Karlicky, M., Vrsnak, B., Temmer, M.,
 Magdalenic, J., Dennis, B.R., Otruba, W., Poetzi, W. 2006, \aap, 446, 675}

\bibitem[Watanabe(1990)]{Wata90}{Watanabe, T. 1990, \solphys, 126, 351}

\bibitem[Wulser et al.(1994)]{Wuls94}
{Wulser, J. P., et al. 1994, \apj, 424, 459}

\bibitem[Xu et al.(2004)]{Xu04}
{Xu, Y., Cao, W., Liu, C., Yang, G., Qiu, J., Jing, J., Denker, C., \& Wang, H. 2004, \apj, 607, L131}

\bibitem[Yokoyama \& Shibata(2001)]{Yoko01}{Yokoyama, T., \& Shibata, K. 2001, \apj, 549, 1160}



\end{thebibliography}
\end{document}